\documentclass[11pt]{article}
\usepackage{graphicx}
\usepackage{amsmath,amssymb,amsthm,amsxtra,overpic,bbm,bm,epsfig,ulem}
\usepackage{color}
\usepackage{multirow}
\usepackage{rotating}
\usepackage{multicol}
\usepackage{array}

\usepackage{array}
\newcommand{\PreserveBackslash}[1]{\let\temp=\\#1\let\\=\temp}
\newcolumntype{C}[1]{>{\PreserveBackslash\centering}p{#1}}
\newcolumntype{R}[1]{>{\PreserveBackslash\raggedleft}p{#1}}
\newcolumntype{L}[1]{>{\PreserveBackslash\raggedright}p{#1}}

\makeatletter

\newcommand{\Rmnum}[1]{\expandafter\@slowromancap\romannumeral #1@}
\makeatother

\begin{document}

\begin{center}
{\Large \bf Further study on the textures of neutrino mass matrix  \\
for maximal atmospherical mixing angle and Dirac CP phase}
\end{center}

\vspace{0.05cm}

\begin{center}
{\bf Zhi-Cheng Liu, Chong-Xing Yue and Zhen-hua Zhao \footnote{E-mail: zhzhao@itp.ac.cn}  } \\
{ Department of Physics, Liaoning Normal University, Dalian 116029, China }
\end{center}

\vspace{0.2cm}

\begin{abstract}
In this paper, we derive in a novel approach the possible textures of neutrino mass matrix that can lead to maximal atmospherical mixing angle ($\theta^{}_{23} = \pi/4$) and Dirac CP phase ($\delta = - \pi/2$) in two phenomenologically appealing scenarios: (1) one neutrino mass matrix element being vanishing (2) one neutrino mass being vanishing. For the obtained textures, some neutrino mass sum rules which relate the neutrino masses and mixing parameters will emerge. With the help of these sum rules, the unknown absolute neutrino mass scale and Majorana CP phases can be determined. Some discussions about the possible textures of neutrino mass matrix that can lead to $\theta^{}_{23} = \pi/4$, $\delta = - \pi/2$ and maximal Majorana CP phases ($\rho, \sigma = \pi/4$ or $3\pi/4$) as well as the model realization and breakings of the obtained textures are also given.
\end{abstract}

\newpage

\section{Introduction}

Thanks to the various neutrino oscillation experiments, it has been established that neutrinos have small but non-vanishing masses and mix among different flavors \cite{pdg}. On the one hand, the smallness of neutrino masses $m^{}_{i}$ (for $i=1, 2, 3$) can be naturally explained by the seesaw mechanism \cite{seesaw}. And the neutrino masses generated via this mechanism are of the Majorana nature in most cases \footnote{Noteworthy, the neutrino masses resulting from the seesaw mechanism can be of the Dirac nature in some special cases \cite{Dirac}.}
for which the mass matrix $M^{}_\nu$ is a complex symmetric one. On the other hand, the neutrino mixing matrix is given by $U= U^{\dagger}_l U^{}_\nu$ where $U^{}_l$ and $U^{}_\nu$ respectively result from diagonalization of the charged lepton mass matrix $M^{}_l$ and $M^{}_\nu$ \cite{pmns}. In the commonly used basis of $M^{}_l$ being diagonal which is also adopted here, $U$ can be identified with the unitary matrix (i.e., $U^{}_\nu$) for diagonalizing $M^{}_\nu$:
\begin{eqnarray}
U^\dagger M^{}_\nu U^* = {\rm Diag}\left( m^{}_1, m^{}_2, m^{}_3 \right) \;.
\label{1}
\end{eqnarray}
In the standard way, it is parameterized as
\begin{eqnarray}
U = P^{}_l O^{}_{23} U^{}_{13} O^{}_{12} P^{}_\nu \;,
\label{2}
\end{eqnarray}
where $P^{}_l = {\rm Diag} \left(e^{{\rm i} \phi_e},  e^{{\rm i} \phi_\mu},  e^{{\rm i} \phi_\tau} \right)$ and $P^{}_\nu = {\rm Diag} \left(e^{{\rm i} \rho},  e^{{\rm i} \sigma}, 1 \right)$ are two diagonal phase matrices, and
\begin{eqnarray}
O^{}_{23} = \begin{pmatrix}
1 & 0 & 0  \\ 0 & c^{}_{23} & s^{}_{23} \\ 0 & - s^{}_{23} & c^{}_{23}
\end{pmatrix} \;, \hspace{1cm}
U^{}_{13} = \begin{pmatrix}
c^{}_{13} & 0 & s^{}_{13} e^{-{\rm i}\delta} \\ 0 & 1 & 0 \\ - s^{}_{13} e^{{\rm i}\delta} & 0 & c^{}_{13}
\end{pmatrix} \;, \hspace{1cm}
O^{}_{12} = \begin{pmatrix}
c^{}_{12} & s^{}_{12} & 0 \\ - s^{}_{12} & c^{}_{12} & 0 \\ 0 & 0 & 1
\end{pmatrix} \;,
\label{3}
\end{eqnarray}
with $c^{}_{ij} = \cos{\theta^{}_{ij}}$ and $s^{}_{ij} = \sin{\theta^{}_{ij}}$ for the mixing angles $\theta^{}_{ij}$ (for $ij = 12, 13, 23$). As for the phases, $\delta$ is ($\rho$ and $\sigma$ are) the Dirac (Majorana) CP phase(s), while $\phi^{}_{e, \mu, \tau}$ are unphysical phases that can be removed by the redefinitions of charged lepton fields. In addition, neutrino oscillations are also governed by two independent neutrino mass squared differences $\Delta m^{2}_{ij} =  m^2_i - m^2_j$ (for $ij = 21, 31$).

So far, the neutrino oscillation experiments give us the following results for the neutrino mass squared differences \cite{global}
\begin{eqnarray}
\Delta m^2_{21} = \left(7.50^{+0.19}_{-0.17}\right)\times 10^{-5} \ {\rm eV^2}  \;, \hspace{1cm}
|\Delta m^2_{31}| = (2.524^{+0.039}_{-0.040})\times 10^{-3} \ {\rm eV^2} \;.
\label{4}
\end{eqnarray}
Note that the sign of $\Delta m^2_{31}$ has not yet been determined, thereby allowing for two possible neutrino mass orderings: $m^{}_1 < m^{}_2 < m^{}_3$ (the normal neutrino mass ordering and NO for short) and $m^{}_3 < m^{}_1 < m^{}_2$ (the inverted neutrino mass ordering and IO for short). On the other side, the mixing parameters $\theta^{}_{13}$, $\theta^{}_{23}$ and $\delta$ take the values of
\begin{eqnarray}
\sin^2{\theta^{}_{13}} = 0.02166 \pm 0.00075  \;, \hspace{1cm}
\sin^2{\theta^{}_{23}} = 0.441 \pm 0.024 \;, \hspace{1cm}
\delta = 261^\circ \pm 55^\circ \;,
\label{5}
\end{eqnarray}
in the NO case, or
\begin{eqnarray}
\sin^2{\theta^{}_{13}} = 0.02179 \pm 0.00076  \;, \hspace{1cm}
\sin^2{\theta^{}_{23}} = 0.587 \pm 0.022 \;, \hspace{1cm}
\delta = 277^\circ \pm 43^\circ \;,
\label{6}
\end{eqnarray}
in the IO case, while $\theta^{}_{12}$ takes the value of $\sin^2{\theta^{}_{12}} = 0.306 \pm 0.012$
in either mass ordering case \cite{global}.

However, the neutrino oscillation experiments are insensitive to the absolute neutrino mass scale and Majorana CP phases. Information about them can only be inferred from non-oscillatory experiments: (1) The beta decay experiments can probe the effective electron neutrino mass $m^{}_\beta = \sqrt{\sum m^{}_i |U^{}_{ei}|^2}$ (with $U^{}_{ei}$ being the $i$-th element in the first row of $U$) by measuring the endpoint of the spectrum of electrons in beta decays. The current upper limit for it is around 2 eV \cite{mb}, while the future KATRIN experiment is expected to achieve a sensitivity of 0.2 eV at 90\% C.L. \cite{KATRIN}.
(2) The cosmological measurements can probe the sum of neutrino masses $\Sigma = \sum m^{}_i$ by virtue of its effects on cosmic structure formation \cite{cosmos}. And the current upper limit for it is 0.12 eV \cite{Planck}. But it should be noted that the neutrino mass limit obtained from the cosmological measurements is strongly dependent on the cosmological model and observation data used. (3) The lepton number violating (LNV) processes can even directly probe the magnitudes of neutrino mass matrix elements, as an LNV process with the charged leptons $\alpha$ and $\beta$ (for $\alpha, \beta =e, \mu, \tau$) in the final state is governed by $|M^{}_{\alpha \beta}|$ (with $M^{}_{\alpha \beta}$ being the $\alpha\beta$ element of $M^{}_\nu$). At present, the neutrino-less double beta decay, which is governed by $|M^{}_{ee}|$ \cite{0nbb}, is the only feasible process to probe LNV. The current upper limit for $|M^{}_{ee}|$ is 0.2$-$0.4 eV, where the large uncertainty is due to the inconclusive nuclear physics calculations \cite{KGE}.

On the theoretical side, one of the most important goals of neutrino physics research is to identify the flavor structure of $M^{}_\nu$ and its origin \cite{whitepaper}.
Because of the particular observed neutrino mixing pattern (some parameters of which, as explained soon, are close to certain special values), it is widely expected that $M^{}_\nu$ probably has a special texture which may originate from some underlying flavor theory (especially flavor symmetry \cite{review}). The first step to a convincing flavor theory is to reconstruct $M^{}_\nu$ with the help of existing experimental results \cite{reconstruction}. However, since the absolute neutrino mass scale and Majorana CP phases remain unknown, the existing experimental results are not sufficient for reconstructing $M^{}_\nu$ completely, leaving us with a large variety of possible forms for it. Guided by the principle of using as fewest parameters as possible to explain the observed neutrino physics \cite{occam}, a number of approaches to restricting the form of $M^{}_\nu$ and reducing the number of free parameters have been adopted in the literature, among which the impositions of some vanishing neutrino mass matrix element(s) \cite{texture-zero} or one vanishing neutrino mass \cite{mss} are two very popular ones. These two scenarios are both well motivated from the theoretical point of view and highly predictive from the phenomenological point of view:
(1) The vanishing of neutrino mass matrix element(s) can naturally find an origin from the Abelian flavor symmetries \cite{abelian}, while one neutrino mass necessarily vanishes if only two right-handed neutrinos take effect in the popular type-I seesaw mechanism (the so-called minimal seesaw \cite{mss}). (2) The vanishing of neutrino mass matrix element(s) would result in some testable relations between the neutrino masses and mixing parameters \cite{texture-zero}, while the vanishing of one neutrino mass would additionally lead one Majorana CP phase to be ineffective \cite{mss}. Furthermore, the vanishing of a neutrino mass matrix element (e.g., $M^{}_{ee}$) would render the associated LNV process (e.g., the neutrino-less double beta decay) impotent.
If any of these interesting consequences turns out to be favored by the future measurements, the corresponding specific neutrino mass matrix texture will stand out and shed some light on the underlying flavor theory.
Therefore, it makes a lot of sense to phenomenologically study the possible neutrino mass matrix textures featuring vanishing elements or mass eigenvalue.

It is interesting to note that the current neutrino oscillation data is consistent with maximal atmospherical mixing angle ($\theta^{}_{23} = \pi/4$) and Dirac CP phase ($\delta = - \pi/2$ \footnote{It should be noted that the current measurement error for $\delta$ is quite large. So we still need a precise measurement for it to see if it is really close to $-\pi/2$. It is expected that this problem will be cleared up over the next few years by the long baseline neutrino oscillation experiments \cite{LBL}.}). As mentioned above, these remarkable values may point towards some special texture of $M^{}_{\nu}$. In this connection, the specific texture given by the $\mu$-$\tau$ reflection symmetry \cite{MTR} serves as a unique example. This symmetry is defined as follows: $M^{}_\nu$ keeps invariant under a combination of the $\mu$-$\tau$ interchange and CP conjugate operations
\begin{eqnarray}
\nu^{}_e \leftrightarrow \nu^c_e \;, \hspace{1cm} \nu^{}_\mu \leftrightarrow \nu^c_\tau \;,
\hspace{1cm} \nu^{}_\tau \leftrightarrow \nu^c_\mu \;,
\label{7}
\end{eqnarray}
and is characterized by
\begin{eqnarray}
M^{}_{e\mu} = M^*_{e\tau} \;, \hspace{1cm} M^{}_{\mu\mu} = M^*_{\tau\tau}  \;, \hspace{1cm}
M^{}_{ee} \ {\rm and} \ M^{}_{\mu\tau} \ {\rm being \ real}  \;.
\label{8}
\end{eqnarray}
Such a texture leads to the following predictions for the neutrino mixing parameters \cite{GL}
\begin{eqnarray}
\phi^{}_e = \frac{\pi}{2} \;, \hspace{1cm} \phi^{}_\mu = - \phi^{}_\tau \;, \hspace{1cm} \theta^{}_{23} = \frac{\pi}{4} \;, \hspace{1cm} \delta = \pm \frac{\pi}{2} \;, \hspace{1cm} \rho, \sigma = 0 \ {\rm or} \ \frac{\pi}{2} \;.
\label{9}
\end{eqnarray}
However, the $\mu$-$\tau$ reflection symmetry is over restrictive in the sense that its predictions for trivial Majorana CP phases (i.e., $\rho, \sigma = 0$ or $\pi/2$) are not promised by the experimental results. Although this symmetry deserves a particular attention due to its interesting properties, it is reasonable to have an open mind for other possible values of $\rho$ and $\sigma$. For this consideration, in a previous work \cite{LYZ} we did an exercise to derive the possible textures of neutrino mass matrix that can lead to $\theta^{}_{23} = \pi/4$ and $\delta = - \pi/2$ \cite{related} where $\rho$ and $\sigma$ are allowed to take values other than 0 and $\pi/2$.

In this work we attempt to derive the possible textures of neutrino mass matrix that can lead to $\theta^{}_{23} = \pi/4$ and $\delta = - \pi/2$ in two phenomenologically appealing scenarios: (1) one neutrino mass matrix element being vanishing (2) one neutrino mass being vanishing.
The rest part of this paper is organized as follows. In section 2, we briefly recapitulate our approach and some useful results developed in the section 2 and subsections 3.1$-$3.4 of Ref. \cite{LYZ} as a basis for the study performed here. The studies in the scenarios of one neutrino mass matrix element and one neutrino mass being vanishing are carried out in sections 3 and 4, respectively. Some discussions about the possible textures of neutrino mass matrix that can lead to $\theta^{}_{23} = \pi/4$, $\delta = - \pi/2$ and maximal Majorana CP phases (i.e., $\rho, \sigma = \pi/4$ or $3\pi/4$) as well as the model realization and breakings of the obtained textures are given in section 5. Finally, in section 6 we summarize our main results.

\section{The approach and basis results}

In order to avoid the uncertainties due to the unphysical phases, we choose to work on $\bar M^{}_\nu = P^\dagger_l M^{}_\nu P^*_l$ instead of $M^{}_\nu$. One can recover the results for $M^{}_\nu$ from those for $\bar M^{}_\nu$ by simply making the replacements $\bar M^{}_{\alpha \beta} = M^{}_{\alpha \beta} e^{-{\rm i}(\phi^{}_\alpha+ \phi^{}_\beta)}$ (with $\bar M^{}_{\alpha \beta}$ being the $\alpha \beta$ element of $\bar M^{}_\nu$). By definition, $\bar M^{}_\nu$ can be diagonalized in a way as
\begin{eqnarray}
O^{T}_{12} U^\dagger_{13} O^{T}_{23} \bar M^{}_\nu O^{}_{23} U^*_{13} O^{}_{12} = {\rm Diag}\left( m^{}_1 e^{2{\rm i}\rho}, m^{}_2e^{2{\rm i}\sigma}, m^{}_3 \right) \;.
\label{10}
\end{eqnarray}
In light of the purpose of this study, we take $\theta^{}_{23} = \pi/4$ and $\delta = - \pi/2$ in $O^{}_{23}$ and $U^{}_{13}$.
To simplify the expressions in the following discussions, we define the following three matrices in order
\begin{eqnarray}
M^{}_{\rm X}  = O^{T}_{23} \bar M^{}_\nu O^{}_{23} \;, \hspace{1cm}
M^{}_{\rm Y} =  U^{\dagger}_{13} M^{}_{\rm X} U^{*}_{13} \;, \hspace{1cm}
M^{}_{\rm Z}  =  O^{T}_{12}  M^{}_{\rm Y} O^{}_{12} \;,
\label{11}
\end{eqnarray}
whose elements appear as
\begin{eqnarray}
&& M^{}_{\rm X 11} = \bar M^{}_{ee} \;, \hspace{0.5cm} M^{}_{\rm X 12} = \displaystyle \frac{\bar M^{}_{e \mu} - \bar M^{}_{e \tau}}{\sqrt 2} \;, \hspace{0.5cm}  M^{}_{\rm X 13} = \displaystyle \frac{\bar M^{}_{e \mu} + \bar M^{}_{e \tau}}{\sqrt 2} \;,  \nonumber \\
&& M^{}_{\rm X 22} = \displaystyle \frac{\bar M^{}_{\mu \mu} + \bar M^{}_{\tau \tau}}{2} - \bar M^{}_{\mu \tau} \;, \hspace{0.5cm} M^{}_{\rm X23} = \displaystyle \frac{\bar M^{}_{\mu \mu} - \bar M^{}_{\tau \tau}}{2} \;, \hspace{0.5cm}
M^{}_{\rm X 33} = \displaystyle \frac{\bar M^{}_{\mu \mu} + \bar M^{}_{\tau \tau}}{2} + \bar M^{}_{\mu \tau} \;, \nonumber \\
&& M^{}_{\rm Y 11} = c^2_{13} M^{}_{\rm X 11} - {\rm i} \sin 2 \theta^{}_{13} M^{}_{\rm X 13} - s^2_{13} M^{}_{\rm X 33} \;, \hspace{0.5cm}
M^{}_{\rm Y 12} = c^{}_{13} M^{}_{\rm X 12} - {\rm i} s^{}_{13} M^{}_{\rm X 23} \;, \nonumber \\
&& M^{}_{\rm Y 13} = \cos 2 \theta^{}_{13} M^{}_{\rm X 13} - \displaystyle \frac{{\rm i}}{2} \sin 2 \theta^{}_{13} \left( M^{}_{\rm X 11} + M^{}_{\rm X 33} \right) \;, \hspace{0.5cm} M^{}_{\rm Y 22} = M^{}_{\rm X 22} \;, \nonumber \\
&& M^{}_{\rm Y 23} = c^{}_{13} M^{}_{\rm X 23} - {\rm i} s^{}_{13} M^{}_{\rm X 12} \;, \hspace{0.5cm}
M^{}_{\rm Y 33} = c^2_{13} M^{}_{\rm X 33} - {\rm i} \sin 2 \theta^{}_{13} M^{}_{\rm X 13} - s^2_{13} M^{}_{\rm X 11} \;, \nonumber \\
&& M^{}_{\rm Z 11} = c^2_{12} M^{}_{\rm Y 11} - \sin 2 \theta^{}_{12} M^{}_{\rm Y 12} + s^2_{12} M^{}_{\rm Y 22} \;, \hspace{0.5cm}  M^{}_{\rm Z 33} = M^{}_{\rm Y 33} \;, \nonumber \\
&& M^{}_{\rm Z12} = \cos 2 \theta^{}_{12} M^{}_{\rm Y 12} + \displaystyle \frac{1}{2} \sin 2 \theta^{}_{12} \left( M^{}_{\rm Y 11} - M^{}_{\rm Y 22} \right) \;, \hspace{0.5cm} M^{}_{\rm Z 13} = c^{}_{12} M^{}_{\rm Y 13} - s^{}_{12} M^{}_{\rm Y 23} \;,  \nonumber \\ && M^{}_{\rm Z 22} = s^2_{12} M^{}_{\rm Y 11} + \sin 2 \theta^{}_{12} M^{}_{\rm Y 12} + c^2_{12} M^{}_{\rm Y 22} \;,  \hspace{0.5cm}  M^{}_{\rm Z 23} = s^{}_{12} M^{}_{\rm Y 13} + c^{}_{12} M^{}_{\rm Y 23}  \;,
\label{12}
\end{eqnarray}
with $M^{}_{\rm X11}$ being the $11$ element of $M^{}_{\rm X}$ and so on.
By comparing the two sides of Eq. (\ref{10}), one gets the following seven diagonalization conditions (which are referred to as A$-$G in order)
\begin{eqnarray}
& {\rm A:} \quad {\rm Re} \left( M^{}_{\rm Y 13} \right) = 0 & \quad  \Longrightarrow  \quad 2 \cos 2 \theta^{}_{13} R^{}_{\rm X 13} = - \sin 2 \theta^{}_{13} \left( I^{}_{\rm X11} + I^{}_{\rm X 33} \right) \;, \nonumber \\
& {\rm B:} \quad {\rm Im} \left( M^{}_{\rm Y 13} \right) = 0 & \quad \Longrightarrow  \quad 2 \cos 2 \theta^{}_{13} I^{}_{\rm X 13} =  \sin 2 \theta^{}_{13} \left( R^{}_{\rm X11} + R^{}_{\rm X33} \right) \;, \nonumber \\
& {\rm C:} \quad {\rm Re} \left( M^{}_{\rm Y 23} \right) =0 & \quad \Longrightarrow  \quad c^{}_{13} R^{}_{\rm X23} = - s^{}_{13} I^{}_{\rm X12} \;, \nonumber \\
& {\rm D:} \quad {\rm Im} \left( M^{}_{\rm Y 23} \right) =0 & \quad \Longrightarrow  \quad c^{}_{13} I^{}_{\rm X23} =   s^{}_{13} R^{}_{\rm X12} \;, \nonumber \\
& {\rm E:} \quad {\rm Re} \left( M^{}_{\rm Z 12} \right) =0 & \quad \Longrightarrow  \quad 2 \cos 2 \theta^{}_{12} R^{}_{\rm Y12} = - \sin 2 \theta^{}_{12} \left( R^{}_{\rm Y11} - R^{}_{\rm Y22} \right) \;, \nonumber \\
& {\rm F:} \quad {\rm Im} \left( M^{}_{\rm Z 12} \right) =0 & \quad \Longrightarrow  \quad 2 \cos 2 \theta^{}_{12} I^{}_{\rm Y12} = - \sin 2 \theta^{}_{12} \left( I^{}_{\rm Y11} - I^{}_{\rm Y22} \right) \;, \nonumber \\
& {\rm G:} \quad {\rm Im} \left( M^{}_{\rm Z 33} \right)  = 0 & \quad \Longrightarrow  \quad \sin 2 \theta^{}_{13} R^{}_{\rm X13} = c^2_{13} I^{}_{\rm X33} - s^2_{13} I^{}_{\rm X11} \;,
\label{15}
\end{eqnarray}
with $R^{}_{\rm X13} = {\rm Re} \left( M^{}_{\rm X13} \right)$, $I^{}_{\rm X11} = {\rm Im} \left( M^{}_{\rm X11} \right)$ and so on, and
\begin{eqnarray}
M^{}_{\rm Z 11} = m^{}_1 e^{2{\rm i}\rho} \;, \hspace{1cm}   M^{}_{\rm Z 22} = m^{}_2 e^{2{\rm i}\sigma} \;, \hspace{1cm}  {\rm Re}\left( M^{}_{\rm Z 33} \right) = m^{}_3 \;.
\label{14}
\end{eqnarray}

Because we only have two parameters (i.e., $\theta^{}_{12}$ and $\theta^{}_{13}$) at hand to diagonalize $\bar M^{}_\nu$ by making the seven diagonalization conditions hold, there exist five constraint equations for $\bar R^{}_{\alpha \beta} = {\rm Re} ( \bar M^{}_{\alpha \beta} ) $ and $\bar I^{}_{\alpha \beta} = {\rm Im} ( \bar M^{}_{\alpha \beta} )$:
By relating the expressions for $\theta^{}_{13}$ derived from diagonalization conditions A$-$D, one obtains the following constraint equations
\begin{eqnarray}
&& {\rm AB}: \quad \left(\bar R^{}_{e \mu} + \bar R^{}_{e \tau}\right) \left(\bar R^{}_{ee} + \bar R^{}_{\mu \tau} + \frac{\bar R^{}_{\mu \mu} + \bar R^{}_{\tau \tau}}{2} \right)
= - \left(\bar I^{}_{e \mu} + \bar I^{}_{e \tau}\right) \left(\bar I^{}_{ee} + \bar I^{}_{\mu \tau} + \frac{\bar I^{}_{\mu \mu} + \bar I^{}_{\tau \tau}}{2} \right) \;, \nonumber \\
&& {\rm AC}: \quad \left(\bar I^{}_{e \mu} - \bar I^{}_{e \tau}\right) \left(\bar R^{}_{\mu \mu} - \bar R^{}_{\tau \tau}\right)
\left(\bar I^{}_{ee} + \bar I^{}_{\mu \tau} + \frac{\bar I^{}_{\mu \mu} + \bar I^{}_{\tau \tau}}{2} \right)
= \left(\bar R^{}_{e \mu} + \bar R^{}_{e \tau}\right) \Big[ \left(\bar I^{}_{e \mu} - \bar I^{}_{e \tau}\right)^2  \nonumber \\
&& \hspace{10.cm} - \frac{1}{2} \left(\bar R^{}_{\mu \mu} - \bar R^{}_{\tau \tau}\right)^2 \Big]
\;, \nonumber \\
&& {\rm AD}: \quad \left( \bar I^{}_{\mu \mu} - \bar I^{}_{\tau \tau} \right) \left(\bar R^{}_{e \mu} - \bar R^{}_{e \tau}\right)
\left(\bar I^{}_{ee} + \bar I^{}_{\mu \tau} + \frac{\bar I^{}_{\mu \mu} + \bar I^{}_{\tau \tau}}{2} \right)
= - \left(\bar R^{}_{e \mu} + \bar R^{}_{e \tau}\right) \Big[ \left( \bar R^{}_{e \mu} - \bar R^{}_{e \tau} \right)^2 \nonumber \\
&& \hspace{10cm} - \frac{1}{2} \left( \bar I^{}_{\mu \mu} - \bar I^{}_{\tau \tau} \right)^2 \Big] \;, \nonumber \\
&& {\rm BC}: \quad \left(\bar I^{}_{e \mu} - \bar I^{}_{e \tau}\right) \left(\bar R^{}_{\mu \mu} - \bar R^{}_{\tau \tau}\right)
\left(\bar R^{}_{ee} + \bar R^{}_{\mu \tau} + \frac{\bar R^{}_{\mu \mu} + \bar R^{}_{\tau \tau}}{2} \right)
= - \left(\bar I^{}_{e \mu} + \bar I^{}_{e \tau}\right) \Big[ \left(\bar I^{}_{e \mu} - \bar I^{}_{e \tau}\right)^2 \nonumber \\
&& \hspace{10.5cm} - \frac{1}{2} \left(\bar R^{}_{\mu \mu} - \bar R^{}_{\tau \tau}\right)^2 \Big]
\;, \nonumber \\
&& {\rm BD}: \quad \left( \bar I^{}_{\mu \mu} - \bar I^{}_{\tau \tau} \right) \left(\bar R^{}_{e \mu} - \bar R^{}_{e \tau}\right)
\left(\bar R^{}_{ee} + \bar R^{}_{\mu \tau} + \frac{\bar R^{}_{\mu \mu} + \bar R^{}_{\tau \tau}}{2} \right)
= \left(\bar I^{}_{e \mu} + \bar I^{}_{e \tau}\right) \Big[ \left( \bar R^{}_{e \mu} - \bar R^{}_{e \tau} \right)^2 \nonumber \\
&& \hspace{10.5cm} - \frac{1}{2} \left( \bar I^{}_{\mu \mu} - \bar I^{}_{\tau \tau} \right)^2 \Big]
\;, \nonumber \\
&& {\rm CD}: \quad \left(\bar R^{}_{\mu \mu} - \bar R^{}_{\tau \tau}\right) \left(\bar R^{}_{e \mu} - \bar R^{}_{e \tau}\right)
= - \left(\bar I^{}_{\mu \mu} - \bar I^{}_{\tau \tau}\right) \left(\bar I^{}_{e \mu} - \bar I^{}_{e \tau}\right) \;,
\label{16}
\end{eqnarray}
where the symbol AB (and so on) is used to indicate that the referred constraint equation results from a combination of diagonalization conditions A and B (and so on).
It is easy to see that only three of these six constraint equations are independent ones.
By relating the expressions for $\theta^{}_{12}$ derived from diagonalization conditions E and F, one arrives at the constraint equation
\begin{eqnarray}
&& {\rm EF}: \quad R^{}_{\rm Y 12} \left( I^{}_{\rm Y 11} - I^{}_{\rm Y22} \right) = I^{}_{\rm Y12} \left( R^{}_{\rm Y11} - R^{}_{\rm Y22} \right) \;,
\label{17}
\end{eqnarray}
with
\begin{eqnarray}
R^{}_{\rm Y12}  &=& {\rm sgn} \left(\bar R^{}_{e \mu} - \bar R^{}_{e \tau}\right) \sqrt{\frac{1}{2} \left(\bar R^{}_{e \mu} - \bar R^{}_{e \tau}\right)^2 + \frac{1}{4} \left( \bar I^{}_{\mu \mu} - \bar I^{}_{\tau \tau} \right)^2 } \;, \nonumber  \\
I^{}_{\rm Y12}
&=& {\rm sgn} \left(\bar I^{}_{e \mu} - \bar I^{}_{e \tau}\right) \sqrt{\frac{1}{2} \left(\bar I^{}_{e \mu} - \bar I^{}_{e \tau}\right)^2 + \frac{1}{4} \left( \bar R^{}_{\mu \mu} - \bar R^{}_{\tau \tau} \right)^2 } \;, \nonumber \\
I^{}_{\rm Y11} - I^{}_{\rm Y22} & = & \bar I^{}_{ee} - \bar I^{}_{\mu\mu} - \bar I^{}_{\tau\tau} \;, \nonumber \\
R^{}_{\rm Y11} - R^{}_{\rm Y22}  &=& \frac{\bar R^{}_{ee} + \bar R^{}_{\mu \tau}}{2} - \frac{3}{4} \left(\bar R^{}_{\mu \mu} + \bar R^{}_{\tau \tau}\right) + {\rm sgn} \left(\bar I^{}_{e \mu} + \bar I^{}_{e \tau}\right) \nonumber \\
&& \times \sqrt{\frac{1}{2} \left( \bar I^{}_{e \mu} + \bar I^{}_{e \tau}\right)^2 + \frac{1}{4} \left(\bar R^{}_{ee} + \bar R^{}_{\mu \tau} + \frac{ \bar R^{}_{\mu \mu} + \bar R^{}_{\tau \tau}}{2}\right)^2 } \;.
\label{18}
\end{eqnarray}
Finally, a combination of diagonalization conditions A and G yields
\begin{eqnarray}
&& \hspace{-0.8cm} {\rm AG}: \quad  \bar I^{}_{\mu \tau} - \bar I^{}_{ee} + \frac{\bar I^{}_{\mu \mu} + \bar I^{}_{\tau \tau}}{2}  ={\rm sgn} \left(\bar R^{}_{e \mu} + \bar R^{}_{e \tau}\right) \sqrt{2 \left(\bar R^{}_{e \mu} + \bar R^{}_{e \tau}\right)^2 +  \left(\bar I^{}_{ee} + \bar I^{}_{\mu \tau} + \frac{\bar I^{}_{\mu \mu} + \bar I^{}_{\tau \tau}}{2}  \right)^2} \;.
\label{19}
\end{eqnarray}

It should be noted that $\bar M^{}_\nu$ might have such a special texture that some diagonalization condition(s) can hold automatically independent of the value of $\theta^{}_{12}$ or $\theta^{}_{13}$. (For example, when one has $I^{}_{\rm X 13} = R^{}_{\rm X 11} + R^{}_{\rm X 33} = 0$, diagonalization condition B will hold automatically, in which case we do not need a particular $\theta^{}_{13}$ to make such a diagonalization condition hold.) In total, there are four basic cases of diagonalization condition(s) holding automatically \cite{LYZ}:
(1) Diagonalization conditions A and G simultaneously hold automatically under the conditions of
\begin{eqnarray}
\bar R^{}_{e \mu} = -\bar R^{}_{e \tau} \;, \hspace{1cm} \bar I^{}_{ee} = 0 \;, \hspace{1cm} -2 \bar I^{}_{\mu \tau} = \bar I^{}_{\mu \mu} + \bar I^{}_{\tau \tau} \;,
\label{20}
\end{eqnarray}
which combined with the constraint equations in Eqs. (\ref{16}-\ref{19}) result in a relation between the neutrino masses and mixing parameters (the so-called neutrino mass sum rule \cite{sumrule})
\begin{eqnarray}
m^{}_1 c^2_{12} \sin 2 \rho + m^{}_2 s^2_{12} \sin 2 \sigma = 0 \;.
\label{21}
\end{eqnarray}
(2) Diagonalization condition B holds automatically under the conditions of
\begin{eqnarray}
\bar I^{}_{e \mu} = - \bar I^{}_{e \tau} \;, \hspace{1cm} -2 \left( \bar R^{}_{ee} + \bar R^{}_{\mu \tau} \right) =  \bar R^{}_{\mu \mu} + \bar R^{}_{\tau \tau} \;,
\label{22}
\end{eqnarray}
which combined with the constraint equations in Eqs. (\ref{16}-\ref{19}) result in the neutrino mass sum rule
\begin{eqnarray}
m^{}_1 c^2_{12} \cos 2 \rho + m^{}_2 s^2_{12} \cos 2 \sigma + m^{}_3 = 0 \;.
\label{23}
\end{eqnarray}
(3) Diagonalization conditions C and F simultaneously hold automatically under the conditions of
\begin{eqnarray}
\bar I^{}_{ee} = \bar I^{}_{\mu \mu} + \bar I^{}_{\tau \tau} \;, \hspace{1cm} \bar R^{}_{\mu \mu} = \bar R^{}_{\tau \tau} \;, \hspace{1cm} \bar I^{}_{e \mu} = \bar I^{}_{e \tau}  \;,
\label{24}
\end{eqnarray}
which combined with the constraint equations in Eqs. (\ref{16}-\ref{19}) result in the neutrino mass sum rule
\begin{eqnarray}
m^{}_1 \sin 2 \rho - m^{}_2 \sin 2 \sigma = 0 \;.
\label{25}
\end{eqnarray}
(4) Diagonalization conditions D and E simultaneously hold automatically under the conditions of
\begin{eqnarray}
&& \hspace{-0.7cm} \bar R^{}_{ee} + \bar R^{}_{\mu \tau} - \frac{3}{2} \left(\bar R^{}_{\mu \mu} +\bar  R^{}_{\tau \tau}\right)
= -{\rm sgn} \left(\bar I^{}_{e \mu} + \bar I^{}_{e \tau}\right)
\sqrt{2 \left( \bar I^{}_{e \mu} + \bar I^{}_{e \tau}\right)^2 + \left(\bar R^{}_{ee} + \bar R^{}_{\mu \tau} + \frac{ \bar R^{}_{\mu \mu} +\bar  R^{}_{\tau \tau} } {2}\right)^2 }  \;, \nonumber \\
&& \hspace{-0.7cm} \bar I^{}_{\mu \mu} = \bar I^{}_{\tau \tau} \;, \hspace{1cm} \bar R^{}_{e \mu} =\bar R^{}_{e \tau} \;,
\label{26}
\end{eqnarray}
which combined with the constraint equations in Eqs. (\ref{16}-\ref{19}) result in the neutrino mass sum rule
\begin{eqnarray}
m^{}_1 \cos 2 \rho - m^{}_2 \cos 2 \sigma = 0 \;.
\label{27}
\end{eqnarray}
As pointed out in Ref. \cite{LYZ}, these cases can find a motivation from the partial $\mu$-$\tau$ symmetry \cite{mu-tau, mu-tau2}.
In the previous work \cite{LYZ}, we have studied the various combinations of these cases themselves.
In this work, we will study the various combinations of these cases with two phenomenologically appealing scenarios: (1) one neutrino mass matrix element being vanishing (2) one neutrino mass being vanishing.

\section{One neutrino mass matrix element being vanishing}

In this section, we perform a study on the possible textures of neutrino mass matrix that can lead to
$\theta^{}_{23} = \pi/4$ and $\delta = - \pi/2$ in the scenario of one neutrino mass matrix element being vanishing.
As we will see shortly, the vanishing of a neutrino mass matrix element gives two neutrino mass sum rules.
Therefore, we just need one more neutrino mass sum rule arising from the requirement of some diagonalization condition(s) holding automatically to completely determine the three unknown physical parameters (i.e., the absolute neutrino mass scale and two Majorana CP phases).

For later use, we give the expressions for $\bar M^{}_{\alpha \beta}$ in terms of the neutrino masses and mixing parameters
\begin{eqnarray}
\bar M^{}_{ee} & = &  m^{}_1 e^{2 {\rm i} \rho } c^2_{12} c^2_{13} + m^{}_2 e^{2 {\rm i} \sigma } s^2_{12} c^2_{13} - m^{}_3 s^2_{13} \;, \nonumber \\
\bar M^{}_{e\mu} & = & \frac{1}{\sqrt 2} \left[ m^{}_1 e^{2 {\rm i} \rho } c^{}_{12} \left(- s^{}_{12} + {\rm i} c^{}_{12} s^{}_{13}\right)
+ m^{}_2 e^{2 {\rm i} \sigma } s^{}_{12} \left(c^{}_{12}+ {\rm i} s^{}_{12} s^{}_{13}\right) + {\rm i} m^{}_3 s^{}_{13} \right] c^{}_{13} \;, \nonumber \\
\bar M^{}_{e\tau} & = & \frac{1}{\sqrt 2} \left[ m^{}_1 e^{2 {\rm i} \rho } c^{}_{12} \left(s^{}_{12}+ {\rm i} c^{}_{12} s^{}_{13}\right)
+ m^{}_2 e^{2 {\rm i} \sigma } s^{}_{12} \left( - c^{}_{12} + {\rm i} s^{}_{12} s^{}_{13}\right) + {\rm i} m^{}_3 s^{}_{13} \right] c^{}_{13} \;, \nonumber \\
\bar M^{}_{\mu\mu} & = & \frac{1}{2} \left[ m^{}_1 e^{2 {\rm i} \rho } \left(s^{}_{12}-{\rm i} c^{}_{12} s^{}_{13}\right)^2
+ m^{}_2 e^{2 {\rm i} \sigma } \left(c^{}_{12}+ {\rm i} s^{}_{12} s^{}_{13}\right)^2 + m^{}_3 c^2_{13}  \right] \;, \nonumber \\
\bar M^{}_{\mu\tau} & = & \frac{1}{2} \left[ - m^{}_1 e^{2 {\rm i} \rho } \left(s^2_{12} + c^2_{12} s^2_{13} \right) - m^{}_2 e^{2 {\rm i} \sigma } \left(c^2_{12}+s^2_{12} s^2_{13}\right) + m^{}_3 c^2_{13}  \right] \;, \nonumber \\
\bar M^{}_{\tau\tau} & = & \frac{1}{2} \left[ m^{}_1 e^{2 {\rm i} \rho } \left(s^{}_{12}+{\rm i} c^{}_{12} s^{}_{13}\right)^2
+ m^{}_2 e^{2 {\rm i} \sigma } \left(c^{}_{12}-{\rm i} s^{}_{12} s^{}_{13}\right)^2 + m^{}_3 c^2_{13} \right] \;,
\label{32}
\end{eqnarray}
which are reconstructed in a way as
\begin{eqnarray}
&& \bar M^{}_\nu = O^{}_{23}  U^{}_{13} O^{}_{12}  {\rm Diag}\left( m^{}_1 e^{2{\rm i}\rho}, m^{}_2e^{2{\rm i}\sigma}, m^{}_3 \right) O^{T}_{12}  U^{T}_{13} O^{T}_{23} \;.
\label{33}
\end{eqnarray}
In the calculations, $\theta^{}_{23} = \pi/4$ and $\delta  = - \pi/2$ have been input.

\subsection{$\bar M^{}_{ee} = 0$}

In the case of $\bar M^{}_{ee} = 0$, one has $I^{}_{\rm X11} = \bar I^{}_{ee} = 0$, which will lead diagonalization conditions A(G) to hold automatically. (Here and in the following we use the phrases ``A(G)", ``C(F)" and ``D(E)" to make it evident that diagonalization conditions A and G, C and F and D and E always simultaneously hold automatically, respectively.) Hence there are four new constraint equations (i.e., those in Eq. (\ref{20}) and $\bar R^{}_{ee} =0$) in addition to those in Eqs. (\ref{16}-\ref{19}). As a result, only three (i.e., EF and two of BC, BD and CD) of the constraint equations in Eqs. (\ref{16}-\ref{19}) are still independent ones. So there are totally seven independent constraint equations, more than those in Eqs. (\ref{16}-\ref{19}) by two in number. It is thus natural to expect that there are two more predictions for the neutrino masses and mixing parameters in addition to $\theta^{}_{23}= \pi/4$ and $\delta = -\pi/2$, which are directly obtained as
\begin{eqnarray}
m^{}_1 c^2_{12} \sin 2 \rho + m^{}_2 s^2_{12} \sin 2 \sigma = 0 \;, \nonumber \\
\left(  m^{}_1 c^2_{12} \cos 2 \rho + m^{}_2 s^2_{12} \cos 2 \sigma \right) c^2_{13} - m^{}_3 s^2_{13} = 0
\;,
\label{34}
\end{eqnarray}
from Eq. (\ref{32}) by taking $\bar M^{}_{ee} =0$. Not surprisingly, one of them is the same as that in Eq. (\ref{21}). In the following, we will study the various cases in which one more neutrino mass sum rule arises from the requirement of some diagonalization condition(s) in addition to A(G) also holding automatically so that the three unknown physical parameters can be completely determined.

In the case of diagonalization condition B holding automatically in combination with $\bar M^{}_{ee} = 0$,  there are the following six new constraint equations (i.e., those in Eqs. (\ref{20}, \ref{22}) and $\bar R^{}_{ee} =0$) in addition to those in Eqs. (\ref{16}-\ref{19})
\begin{eqnarray}
&& \bar R^{}_{e \mu} = -\bar R^{}_{e \tau} \;, \hspace{1cm} \bar I^{}_{ee} = 0 \;, \hspace{1cm} -2 \bar I^{}_{\mu \tau} = \bar I^{}_{\mu \mu} + \bar I^{}_{\tau \tau} \;, \nonumber \\
&& \bar I^{}_{e \mu} = - \bar I^{}_{e \tau} \;, \hspace{1cm}  \bar R^{}_{ee} =0\;, \hspace{1cm} -2 \bar R^{}_{\mu \tau} =  \bar R^{}_{\mu \mu} + \bar R^{}_{\tau \tau} \;,
\label{36}
\end{eqnarray}
which can be recombined into
\begin{eqnarray}
&& \bar M^{}_{e \mu} = -\bar M^{}_{e \tau} \;,  \hspace{1cm}  -2 \bar M^{}_{\mu \tau} =  \bar M^{}_{\mu \mu} + \bar M^{}_{\tau \tau} \;, \hspace{1cm} \bar M^{}_{ee} =0 \;.
\label{37}
\end{eqnarray}
As a result, only two (i.e., CD and EF) of the constraint equations in Eqs. (\ref{16}-\ref{19}) are still independent ones. So there are totally three neutrino mass sum rules, which are given by Eqs. (\ref{23}, \ref{34}). It is found that these sum rules have no chance to be in agreement with the realistic results. (Note that the sum rule in Eq. (\ref{23}) can only be fulfilled in the IO case while the second one in Eq. (\ref{34}) can only be fulfilled in the NO case.)

In the case of diagonalization conditions C(F) holding automatically in combination with $\bar M^{}_{ee} = 0$, there are the following seven new constraint equations (i.e., those in Eqs. (\ref{20}, \ref{24}) and $\bar R^{}_{ee} =0$) in addition to those in Eqs. (\ref{16}-\ref{19})
\begin{eqnarray}
&& \bar R^{}_{e \mu} = -\bar R^{}_{e \tau} \;,  \hspace{1cm} \bar I^{}_{e \mu} = \bar I^{}_{e \tau} \;, \hspace{1cm} \bar R^{}_{\mu \mu} = \bar R^{}_{\tau \tau} \;, \hspace{1cm} \bar I^{}_{\mu \mu} = - \bar I^{}_{\tau \tau} \;, \nonumber \\
&& \bar R^{}_{ee} =\bar I^{}_{ee} = \bar I^{}_{\mu \tau} = 0 \;,
\label{38}
\end{eqnarray}
which can be recombined into
\begin{eqnarray}
\bar M^{}_{e\mu} = - \bar M^*_{e\tau} \;, \hspace{1cm} \bar M^{}_{\mu\mu} = \bar M^*_{\tau\tau}  \;, \hspace{1cm}
\bar M^{}_{ee} =0 \;, \hspace{1cm} \bar M^{}_{\mu\tau} \ {\rm being \ real}  \;.
\label{39}
\end{eqnarray}
As a result, only one (i.e., BD) of the constraint equations in Eqs. (\ref{16}-\ref{19}) is still an independent one. So there are totally three neutrino mass sum rules, which are given by Eqs. (\ref{25}, \ref{34}). By solving these equations, one obtains $m^{}_1 = 0.006$ eV with $[\rho, \sigma] = [0, \pi/2]$ or $0.002$ eV with $[\rho, \sigma] = [\pi/2, 0]$ in the NO case. For these two possible results, the effective electron neutrino mass $m^{}_\beta$ takes a value of 0.011 or 0.010 eV while the neutrino mass sum $\Sigma$ takes a value of 0.068 or 0.062 eV. On the other hand, $\bar M^{}_\nu$ and the magnitudes of its elements (e.g., $\bar M^{}_{ee}$) that govern the associated LNV processes (e.g., the neutrino-less double beta decay) are given by
\begin{eqnarray}
\frac{\bar M^{}_{\nu}}{{\rm eV}} \simeq \begin{pmatrix}
 0 & -0.006-0.005 {\rm i} & 0.006-0.005 {\rm i} \\
\times & 0.022+0.001 {\rm i} & 0.028 \\
\times & \times & 0.022-0.001 {\rm i}
\end{pmatrix} \;, \hspace{1cm} \frac{|\bar M^{}_{\nu}|}{{\rm eV}} \simeq  \begin{pmatrix}
 0 & 0.008 & 0.008 \\
\times & 0.022 & 0.028 \\
\times & \times & 0.022
\end{pmatrix} \;,
\end{eqnarray}
or
\begin{eqnarray}
\frac{\bar M^{}_{\nu}}{{\rm eV}} \simeq  \begin{pmatrix}
 0 & 0.004-0.005 {\rm i}  & -0.004-0.005 {\rm i}  \\
\times  & 0.027-0.001 {\rm i}  & 0.022 \\
\times  & \times & 0.027+0.001 {\rm i}
\end{pmatrix} \;, \hspace{1cm}
\frac{|\bar M^{}_{\nu}|}{{\rm eV}} \simeq \begin{pmatrix}
 0 & 0.006 & 0.006 \\
\times  & 0.027  & 0.022 \\
\times  & \times & 0.027
\end{pmatrix} \;.
\end{eqnarray}
From these results, one can in a more intuitive way appreciate the texture of $\bar M^{}_\nu$.

In the case of diagonalization conditions D(E) holding automatically in combination with $\bar M^{}_{ee} = 0$, there are the following seven new constraint equations (i.e., those in Eqs. (\ref{20}, \ref{26}) and $\bar R^{}_{ee} =0$)
\begin{eqnarray}
&& \bar R^{}_{\mu \tau} - \frac{3}{2} \left(\bar R^{}_{\mu \mu} +\bar  R^{}_{\tau \tau}\right)
= -{\rm sgn} \left(\bar I^{}_{e \mu} + \bar I^{}_{e \tau}\right)
\sqrt{2 \left( \bar I^{}_{e \mu} + \bar I^{}_{e \tau}\right)^2 + \left(\bar R^{}_{\mu \tau} + \frac{ \bar R^{}_{\mu \mu} +\bar  R^{}_{\tau \tau} } {2}\right)^2 }  \;, \nonumber \\
&& \bar R^{}_{ee} = \bar R^{}_{e \mu} = \bar R^{}_{e \tau} = \bar I^{}_{ee} = 0 \;, \hspace{1cm}   \bar I^{}_{\mu \mu}  = \bar I^{}_{\tau \tau} = - \bar I^{}_{\mu \tau} \;,
\label{40}
\end{eqnarray}
in addition to those in Eqs. (\ref{16}-\ref{19}).
As a result, only one (i.e., BC) of the constraint equations in Eqs. (\ref{16}-\ref{19}) is still an independent one. So there are totally three neutrino mass sum rules, which are given by Eqs. (\ref{27}, \ref{34}). By solving these equations, one obtains $m^{}_1 = 0.004$ eV with $[\rho, \sigma] = [0.79\pi, 0.23 \pi]$ in the NO case.
For such a result, $m^{}_\beta$ and $\Sigma$ respectively take a value of 0.010 eV and  0.065 eV, while $\bar M^{}_\nu$ and the magnitudes of its elements are given by
\begin{eqnarray}
\frac{\bar M^{}_{\nu}}{{\rm eV}} \simeq \begin{pmatrix}
0 & -0.001 {\rm i} & -0.010 {\rm i} \\
\times & 0.026+0.003 {\rm i} & 0.024-0.003 {\rm i} \\
\times & \times & 0.024+0.003 {\rm i}
\end{pmatrix} \;, \hspace{1cm}
\frac{|\bar M^{}_{\nu}|}{{\rm eV}} \simeq \begin{pmatrix}
0 & 0.001 & 0.010  \\
\times & 0.026 & 0.024 \\
\times & \times & 0.024
\end{pmatrix}  \;.
\end{eqnarray}

\subsection{$\bar M^{}_{e\mu} = 0$ or $\bar M^{}_{e\tau} =0$}

In the case of $\bar M^{}_{e\mu}  = 0$ $(\bar M^{}_{e\tau} =0)$, we simply get two new constraint equations $\bar R^{}_{e\mu} = \bar I^{}_{e\mu} = 0$ ($\bar R^{}_{e\tau} = \bar I^{}_{e\tau} = 0$) in addition to those in Eqs. (\ref{16}-\ref{19}). The resulting two neutrino mass sum rules are directly obtained as
\begin{eqnarray}
m^{}_1 c^{2}_{12} s^{}_{13} \cos 2 \rho  - (+) m^{}_1 c^{}_{12} s^{}_{12} \sin 2 \rho
+ m^{}_2 s^{2}_{12} s^{}_{13} \cos 2 \sigma + (-)  m^{}_2 c^{}_{12} s^{}_{12} \sin 2 \sigma + m^{}_3 s^{}_{13} = 0 \;, \nonumber \\
m^{}_1 c^{}_{12} s^{}_{12} \cos 2 \rho  + (- ) m^{}_1 c^{2}_{12} s^{}_{13} \sin 2 \rho  - m^{}_2 c^{}_{12} s^{}_{12} \cos 2 \sigma + (-)  m^{}_2 s^{2}_{12} s^{}_{13} \sin 2 \sigma   = 0 \;,
\label{41}
\end{eqnarray}
from Eq. (\ref{32}) by taking $\bar M^{}_{e\mu} =0$ $(\bar M^{}_{e\tau} =0)$.
In the following, we will study the various cases where one more neutrino mass sum rule arises from the requirement of some diagonalization condition(s) holding automatically so that the three unknown physical parameters can be completely determined.
As a result of $\bar M^{}_{e\mu}  = 0$ $(\bar M^{}_{e\tau} =0)$, one has $R^{}_{\rm X12} = - (+) R^{}_{\rm X13}$ and $I^{}_{\rm X12} = - (+) I^{}_{\rm X13}$, which will lead diagonalization conditions A(G) and D(E) and B and C(F) to simultaneously hold automatically, respectively.

In the case of diagonalization conditions A(G) and D(E) holding automatically in combination with $\bar M^{}_{e\mu}  = 0$, there are the following seven new constraint equations (i.e., those in Eqs. (\ref{20}, \ref{26}) and $\bar I^{}_{e\mu} =0$)
\begin{eqnarray}
&& \bar R^{}_{ee} + \bar R^{}_{\mu \tau} - \frac{3}{2} \left(\bar R^{}_{\mu \mu} +\bar  R^{}_{\tau \tau}\right)
= -{\rm sgn} \left(\bar I^{}_{e \tau}\right)
\sqrt{2 \bar I^{2}_{e \tau} + \left(\bar R^{}_{ee} + \bar R^{}_{\mu \tau} + \frac{ \bar R^{}_{\mu \mu} +\bar  R^{}_{\tau \tau} } {2}\right)^2 }  \;, \nonumber \\
&& \bar R^{}_{e \mu} = \bar R^{}_{e \tau} = \bar I^{}_{ee} =\bar I^{}_{e\mu} = 0 \;, \hspace{1cm}   \bar I^{}_{\mu \mu}  = \bar I^{}_{\tau \tau} = - \bar I^{}_{\mu \tau} \;,
\label{42}
\end{eqnarray}
in addition to those in Eqs. (\ref{16}-\ref{19}).
As a result, only one (i.e., BC) of the constraint equations in Eqs. (\ref{16}-\ref{19}) is still an independent one. So there are
totally three neutrino mass sum rules, which are given by three independent ones of Eqs. (\ref{21}, \ref{27}, \ref{41}). By solving these equations, one obtains $m^{}_1 = 0.009$ eV with $[\rho, \sigma] = [0.41 \pi, 0.65 \pi]$ in the NO case or $m^{}_3 = 0.006$ eV with $[\rho, \sigma] = [0.51 \pi, 0.47 \pi]$ in the IO case.
For these two possible results, $m^{}_\beta$ takes a value of 0.012 or 0.050 eV while $\Sigma$ takes a value of 0.072 or 0.11 eV. On the other hand, $\bar M^{}_\nu$ and the magnitudes of its elements are given by
\begin{eqnarray}
\frac{\bar M^{}_{\nu}}{{\rm eV}} \simeq \begin{pmatrix}
-0.008 & 0 & 0.009 {\rm i} \\
\times & 0.022-0.003 {\rm i} & 0.029+0.003 {\rm i} \\
\times & \times & 0.020-0.003 {\rm i}
\end{pmatrix} \;, \hspace{1cm}
\frac{|\bar M^{}_{\nu}|}{{\rm eV}} \simeq \begin{pmatrix}
0.008 & 0 & 0.009  \\
\times & 0.022 & 0.029 \\
\times & \times & 0.020
\end{pmatrix} \;,
\end{eqnarray}
or
\begin{eqnarray}
\frac{\bar M^{}_{\nu}}{{\rm eV}} \simeq  \begin{pmatrix}
-0.049 & 0 & -0.009  {\rm i} \\
\times & -0.022+0.003  {\rm i} & 0.028-0.003  {\rm i} \\
\times & \times & -0.020+0.003 {\rm i}
\end{pmatrix} \;, \hspace{1cm}
\frac{|\bar M^{}_{\nu}|}{{\rm eV}} \simeq \begin{pmatrix}
0.049 & 0 & 0.009  \\
\times & 0.022 & 0.028 \\
\times & \times & 0.020
\end{pmatrix} \;.
\end{eqnarray}

In the case of diagonalization conditions B and C(F) holding automatically in combination with $\bar M^{}_{e\mu}  = 0$, there are the following six new constraint equations (i.e., those in Eqs. (\ref{22}, \ref{24}) and $\bar R^{}_{e\mu} =0$)
\begin{eqnarray}
\bar I^{}_{ee} = \bar I^{}_{\mu \mu} + \bar I^{}_{\tau \tau} \;, \hspace{1cm} \bar R^{}_{\mu \mu} = \bar R^{}_{\tau \tau} = - \left( \bar R^{}_{ee} + \bar R^{}_{\mu \tau} \right) \;, \hspace{1cm} \bar R^{}_{e\mu} =\bar I^{}_{e \mu} = \bar I^{}_{e \tau} = 0  \;,
\label{44}
\end{eqnarray}
in addition to those in Eqs. (\ref{16}-\ref{19}).
As a result, only two (i.e., AD and AG) of the constraint equations in Eqs. (\ref{16}-\ref{19}) are still independent ones. So there are totally three neutrino mass sum rules, which are given by three independent ones of Eqs. (\ref{23}, \ref{25}, \ref{41}). By solving these equations, one obtains $m^{}_3 = 0.0007$ eV with $[\rho, \sigma] = [0.27 \pi, 0.22 \pi]$ in the IO case.
For such a result, $m^{}_\beta$ and $\Sigma$ respectively take a value of 0.049 eV and 0.10 eV, while $\bar M^{}_\nu$ and the magnitudes of its elements are given by
\begin{eqnarray}
\frac{\bar M^{}_{\nu}}{{\rm eV}} \simeq \begin{pmatrix}
-0.001+0.048 {\rm i} & 0 & -0.010 \\
\times & 0.003+0.025 {\rm i} & -0.002-0.025 {\rm i} \\
\times & \times & 0.003+0.023 {\rm i}
\end{pmatrix} \;, \hspace{0.5cm}
\frac{|\bar M^{}_{\nu}|}{{\rm eV}} \simeq \begin{pmatrix}
0.048 & 0 & 0.010 \\
\times & 0.025  & 0.025  \\
\times & \times & 0.023
\end{pmatrix}  \;.
\end{eqnarray}

In the case of some diagonalization condition(s) holding automatically in combination with $\bar M^{}_{e\tau} = 0$, from the above results in the case of the same diagonalization condition(s) holding automatically in combination with $\bar M^{}_{e\mu} = 0$, the constraint equations and the resulting $\bar M^{}_\nu$ can be obtained by making the interchanges $\bar R^{}_{e\mu} \leftrightarrow - \bar R^{}_{e\tau}$, $\bar I^{}_{e\mu} \leftrightarrow \bar I^{}_{e\tau}$, $\bar R^{}_{\mu\mu} \leftrightarrow  \bar R^{}_{\tau\tau}$ and $\bar I^{}_{\mu\mu} \leftrightarrow  - \bar I^{}_{\tau\tau}$ and a sign change for $\bar I^{}_{ee}$ and $\bar I^{}_{\mu \tau}$, while the predictions for the three unknown physical parameters can be obtained by making the replacements $\rho \to \pi - \rho$ and $\sigma \to \pi -\sigma$. This reflects a symmetry between the $\mu$ and $\tau$ flavors.

\subsection{$\bar M^{}_{\mu \mu} = 0$ or $\bar M^{}_{\tau \tau} = 0$}

In the case of $\bar M^{}_{\mu\mu} = 0$ ($\bar M^{}_{\tau \tau} = 0$), we simply get two new constraint equations $\bar R^{}_{\mu\mu} = \bar I^{}_{\mu\mu} = 0$ ($\bar R^{}_{\tau\tau} = \bar I^{}_{\tau\tau} = 0$) in addition to those in Eqs. (\ref{16}-\ref{19}). The resulting two neutrino mass sum rules are directly obtained as
\begin{eqnarray}
&& m^{}_1  \left( s^{2}_{12} - c^{2}_{12} s^{2}_{13} \right) \cos 2 \rho + (-) 2 m^{}_1 c^{}_{12} s^{}_{12} s^{}_{13} \sin 2 \rho \nonumber \\
&& \hspace{1cm}+ m^{}_2  \left( c^{2}_{12} - s^{2}_{12} s^{2}_{13} \right) \cos 2 \sigma - (+) 2 m^{}_2 c^{}_{12} s^{}_{12} s^{}_{13} \sin 2 \sigma + m^{}_3 c^2_{13} = 0 \;, \nonumber \\
&& 2 m^{}_1 c^{}_{12} s^{}_{12} s^{}_{13} \cos 2 \rho - (+) m^{}_1  \left( s^{2}_{12} - c^{2}_{12} s^{2}_{13} \right) \sin 2 \rho \nonumber \\
&& \hspace{1cm} - 2 m^{}_2 c^{}_{12} s^{}_{12} s^{}_{13} \cos 2 \sigma - (+) m^{}_2  \left( c^{2}_{12} - s^{2}_{12} s^{2}_{13} \right) \sin 2 \sigma  = 0 \;,
\label{45}
\end{eqnarray}
from Eq. (\ref{32}) by taking $\bar M^{}_{\mu\mu} =0$ ($\bar M^{}_{\tau \tau} = 0$). In the following, we will study the various cases where one more neutrino mass sum rule arises from the requirement of some diagonalization condition(s) holding automatically so that the three unknown physical parameters can be completely determined.

In the case of diagonalization conditions A(G) holding automatically in combination with $\bar M^{}_{\mu\mu} = 0$, there are the following five new constraint equations (i.e., those in Eq. (\ref{20}) and $\bar R^{}_{\mu\mu} = \bar I^{}_{\mu\mu} = 0$)
\begin{eqnarray}
\bar R^{}_{e \mu} = -\bar R^{}_{e \tau} \;, \hspace{1cm} \bar I^{}_{ee} = \bar R^{}_{\mu\mu} = \bar I^{}_{\mu\mu} =0 \;, \hspace{1cm} -2 \bar I^{}_{\mu \tau} = \bar I^{}_{\tau \tau} \;,
\label{46}
\end{eqnarray}
in addition to those in Eqs. (\ref{16}-\ref{19}). As a result, only three (i.e., EF and two of BC, BD and CD) of the constraint equations in Eqs. (\ref{16}-\ref{19}) are still independent ones. So there are totally three neutrino mass sum rules, which are given by Eqs. (\ref{21}, \ref{45}). By solving these equations, one obtains $m^{}_3 = 0.024$ eV with $[\rho, \sigma] = [0.97 \pi, 0.43 \pi]$ in the IO case.
For such a result, $m^{}_\beta$ and $\Sigma$ respectively take a value of 0.055 eV and 0.13 eV, while $\bar M^{}_\nu$ and the magnitudes of its elements are given by
\begin{eqnarray}
\frac{\bar M^{}_{\nu}}{{\rm eV}} \simeq \begin{pmatrix}
 0.021 & -0.033+0.016 {\rm i} & 0.033-0.007 {\rm i} \\
\times & 0 & 0.020-0.007 {\rm i} \\
\times & \times & 0.005+0.014 {\rm i}
\end{pmatrix} \;, \hspace{1cm}
\frac{|\bar M^{}_{\nu}|}{{\rm eV}} \simeq \begin{pmatrix}
0.021 & 0.037 & 0.034 \\
\times & 0 & 0.021 \\
\times & \times & 0.015
\end{pmatrix}  \;.
\end{eqnarray}

In the case of diagonalization condition B holding automatically in combination with $\bar M^{}_{\mu\mu} = 0$, there are the following four new constraint equations (i.e., those in Eq. (\ref{22}) and $\bar R^{}_{\mu\mu} = \bar I^{}_{\mu\mu} =0$)
\begin{eqnarray}
\bar I^{}_{e \mu} = - \bar I^{}_{e \tau} \;, \hspace{1cm} -2 \left( \bar R^{}_{ee} + \bar R^{}_{\mu \tau} \right) =   \bar R^{}_{\tau \tau} \;, \hspace{1cm} \bar R^{}_{\mu\mu} = \bar I^{}_{\mu\mu} =0 \;,
\label{47}
\end{eqnarray}
in addition to those in Eqs. (\ref{16}-\ref{19}).
As a result, only four (i.e., AG, EF and two of AC, AD and CD) of the constraint equations in Eqs. (\ref{16}-\ref{19}) are still independent ones. So there are totally three neutrino mass sum rules, which are given by Eqs. (\ref{23}, \ref{45}). By solving these equations, one obtains $m^{}_3 = 0.050$ eV with $[\rho, \sigma] = [0.35 \pi, 0.54 \pi]$ in the IO case.
For such a result, $m^{}_\beta$ and $\Sigma$ respectively take a value of 0.070 eV and 0.19 eV, while $\bar M^{}_\nu$ and the magnitudes of its elements are given by
\begin{eqnarray}
\frac{\bar M^{}_{\nu}}{{\rm eV}} \simeq \begin{pmatrix}
-0.050+0.033 {\rm i} & -0.012-0.024 {\rm i} & 0.005+0.024 {\rm i} \\
\times & 0 & 0.055-0.003 {\rm i} \\
\times & \times & -0.010+0.004 {\rm i}
\end{pmatrix} \;, \hspace{0.5cm}
\frac{|\bar M^{}_{\nu}|}{{\rm eV}} \simeq \begin{pmatrix}
0.060 & 0.027 & 0.025 \\
\times & 0 & 0.055 \\
\times & \times & 0.011
\end{pmatrix}  \;.
\end{eqnarray}

In the case of diagonalization conditions C(F) holding automatically in combination with $\bar M^{}_{\mu\mu} = 0$, there are the following five new constraint equations (i.e., those in Eq. (\ref{24}) and $\bar R^{}_{\mu\mu} = \bar I^{}_{\mu\mu} =0$)
\begin{eqnarray}
\bar I^{}_{ee} = \bar I^{}_{\tau \tau} \;, \hspace{1cm} \bar R^{}_{\mu \mu} = \bar I^{}_{\mu\mu} = \bar R^{}_{\tau \tau} =0 \;, \hspace{1cm} \bar I^{}_{e \mu} = \bar I^{}_{e \tau}  \;,
\label{48}
\end{eqnarray}
in addition to those in Eqs. (\ref{16}-\ref{19}).
As a result, only three (i.e., AG and two of AB, AD and BD) of the constraint equations in Eqs. (\ref{16}-\ref{19}) are still independent ones. So there are totally three neutrino mass sum rules, which are given by Eqs. (\ref{25}, \ref{45}). By solving these equations, one obtains $m^{}_3 = 0.022$ eV with $[\rho, \sigma] = [0.04 \pi, 0.46 \pi]$ in the IO case.
For such a result, $m^{}_\beta$ and $\Sigma$ respectively take a value of 0.055 eV and 0.13 eV, while $\bar M^{}_\nu$ and the magnitudes of its elements are given by
\begin{eqnarray}
\frac{\bar M^{}_{\nu}}{{\rm eV}} \simeq \begin{pmatrix}
0.019+0.014  {\rm i} & -0.035+0.004 {\rm i} & 0.032+0.004  {\rm i} \\
\times & 0 & 0.021-0.007  {\rm i} \\
\times & \times & 0.014  {\rm i}
\end{pmatrix} \;, \hspace{0.5cm}
\frac{|\bar M^{}_{\nu}|}{{\rm eV}} \simeq \begin{pmatrix}
0.024 & 0.035 & 0.032 \\
\times & 0 & 0.022 \\
\times & \times & 0.014
\end{pmatrix}  \;.
\end{eqnarray}

In the case of diagonalization conditions D(E) holding automatically in combination with $\bar M^{}_{\mu\mu} = 0$, there are five new constraint equations (i.e., those in Eq. (\ref{26}) and $\bar R^{}_{\mu\mu} = \bar I^{}_{\mu\mu} =0$) in addition to those in Eqs. (\ref{16}-\ref{19}).
As a result, only three (i.e., AG and two of AB, AC and BC) of the constraint equations in Eqs. (\ref{16}-\ref{19}) are still independent ones. So there are totally three neutrino mass sum rules, which are given by Eqs. (\ref{27}, \ref{45}). It is found that these sum rules have no chance to be in agreement with the realistic results.

In the case of some diagonalization condition(s) holding automatically in combination with $\bar M^{}_{\tau\tau} = 0$, from the above results in the case of the same diagonalization condition(s) holding automatically in combination with $\bar M^{}_{\mu\mu} = 0$, the constraint equations and the resulting $\bar M^{}_\nu$ can be obtained by making the interchanges $\bar R^{}_{e\mu} \leftrightarrow - \bar R^{}_{e\tau}$, $\bar I^{}_{e\mu} \leftrightarrow \bar I^{}_{e\tau}$, $\bar R^{}_{\mu\mu} \leftrightarrow  \bar R^{}_{\tau\tau}$ and $\bar I^{}_{\mu\mu} \leftrightarrow  - \bar I^{}_{\tau\tau}$ and a sign change for $\bar I^{}_{ee}$ and $\bar I^{}_{\mu \tau}$, while the predictions for the three unknown physical parameters can be obtained by making the replacements $\rho \to \pi - \rho$ and $\sigma \to \pi -\sigma$. This reflects a symmetry between the $\mu$ and $\tau$ flavors.

\subsection{$\bar M^{}_{\mu\tau} = 0$}

In the case of $\bar M^{}_{\mu\tau} = 0$, we simply get two new constraint equations $\bar R^{}_{\mu\tau} = \bar I^{}_{\mu\tau} = 0$ in addition to those in Eqs. (\ref{16}-\ref{19}). The resulting two neutrino mass sum rules are directly obtained as
\begin{eqnarray}
&& m^{}_1 \cos 2 \rho \left(s^2_{12} + c^2_{12} s^2_{13}\right) + m^{}_2 \cos 2 \sigma \left(c^2_{12}+s^2_{12} s^2_{13}\right) -m^{}_3 c^2_{13} = 0 \;, \nonumber\\
&& m^{}_1 \sin 2 \rho \left(s^2_{12} + c^2_{12} s^2_{13}\right) + m^{}_2 \sin 2 \sigma \left(c^2_{12}+s^2_{12} s^2_{13}\right) = 0 \;,
\label{49}
\end{eqnarray}
from Eq. (\ref{32}) by taking $\bar M^{}_{\mu\tau} =0$.
In the following, we will study the various cases where one more neutrino mass sum rule arises from the requirement of some diagonalization condition(s) holding automatically so that the three unknown physical parameters can be completely determined. As a result of $\bar M^{}_{\mu\tau} = 0$, one has $I^{}_{\rm Y 11} - I^{}_{\rm Y22} = I^{}_{\rm X11} - 2 I^{}_{\rm X 33}$, which will lead diagonalization conditions A(G) and C(F) to always simultaneously hold automatically.

In the case of diagonalization conditions A(G) and C(F) holding automatically in combination with $\bar M^{}_{\mu\tau} = 0$, there are the following seven new constraint equations (i.e., those in Eqs. (\ref{20}, \ref{24}) and $\bar R^{}_{\mu \tau} =0$) in addition to those in Eqs. (\ref{16}-\ref{19})
\begin{eqnarray}
&& \bar R^{}_{e \mu} = -\bar R^{}_{e \tau} \;,  \hspace{1cm} \bar I^{}_{e \mu} = \bar I^{}_{e \tau} \;, \hspace{1cm} \bar R^{}_{\mu \mu} = \bar R^{}_{\tau \tau} \;, \hspace{1cm} \bar I^{}_{\mu \mu} = - \bar I^{}_{\tau \tau} \;, \nonumber \\
&& \bar I^{}_{ee} = \bar R^{}_{\mu \tau} = \bar I^{}_{\mu \tau} = 0 \;,
\label{50}
\end{eqnarray}
which can be recombined into
\begin{eqnarray}
\bar M^{}_{e\mu} = - \bar M^*_{e\tau} \;, \hspace{1cm} \bar M^{}_{\mu\mu} = \bar M^*_{\tau\tau}  \;, \hspace{1cm}
\bar M^{}_{\mu\tau} =0 \;, \hspace{1cm} \bar M^{}_{ee} \ {\rm being \ real}  \;.
\label{51}
\end{eqnarray}
As a result, only one (i.e., BD) of the constraint equations in Eqs. (\ref{16}-\ref{19}) is still an independent one. So there are totally three neutrino mass sum rules, which are given by three independent ones of Eqs. (\ref{21}, \ref{25}, \ref{49}). By solving these equations, one obtains $m^{}_1 = 0.17$ eV with $[\rho, \sigma] = [0, 0]$ in the NO case
or $m^{}_3 = 0.021$ eV with $[\rho, \sigma] = [\pi/2, 0]$ in the IO case.
For these two possible results, $m^{}_\beta$ takes a value of 0.17 or 0.054 eV while $\Sigma$ takes a value of 0.51 or 0.13 eV. Note that the result of $m^{}_1 = 0.17$ eV is strongly disfavored by the cosmological measurements for the neutrino mass sum.
On the other hand, $\bar M^{}_\nu$ and the magnitudes of its elements are given by
\begin{eqnarray}
\frac{\bar M^{}_{\nu}}{{\rm eV}} \simeq \begin{pmatrix}
0.158 & 0.035 {\rm i} & 0.035 {\rm i} \\
\times & 0.165 & 0 \\
\times & \times & 0.165
\end{pmatrix} \;, \hspace{1cm}
\frac{|\bar M^{}_{\nu}|}{{\rm eV}} \simeq \begin{pmatrix}
0.158 & 0.035  & 0.035  \\
\times & 0.165 & 0 \\
\times & \times & 0.165
\end{pmatrix} \;,
\end{eqnarray}
or
\begin{eqnarray}
\frac{\bar M^{}_{\nu}}{{\rm eV}} \simeq  \begin{pmatrix}
-0.021 & 0.035 & -0.035 \\
\times & 0.021+0.007 {\rm i} & 0 \\
\times & \times & 0.021-0.007{\rm i}
\end{pmatrix} \;, \hspace{1cm}
\frac{|\bar M^{}_{\nu}|}{{\rm eV}} \simeq \begin{pmatrix}
0.021 & 0.035 & 0.035 \\
\times & 0.022 & 0 \\
\times & \times & 0.022
\end{pmatrix} \;.
\end{eqnarray}

In the case of diagonalization condition B holding automatically in combination with $\bar M^{}_{\mu\tau} = 0$, there are four new constraint equations (i.e., those in Eq. (\ref{22}) and $\bar R^{}_{\mu\tau} = \bar I^{}_{\mu\tau} =0$) in addition to those in Eqs. (\ref{16}-\ref{19}). As a result, only four (i.e., AG, EF and two of AC, AD and CD) of the constraint equations in Eqs. (\ref{16}-\ref{19}) are still independent ones. So there are totally three neutrino mass sum rules, which are given by
Eqs. (\ref{23}, \ref{49}). It is found that these sum rules have no chance to be in agreement with the realistic results.

In the case of diagonalization conditions D(E) holding automatically in combination with $\bar M^{}_{\mu\tau} = 0$, there are five new constraint equations (i.e., those in Eq. (\ref{26}) and $\bar R^{}_{\mu\tau} = \bar I^{}_{\mu\tau} =0$) in addition to those in Eqs. (\ref{16}-\ref{19}). As a result, only three (i.e., AG and two of AB, AC and BC) of the constraint equations in Eqs. (\ref{16}-\ref{19}) are still independent ones. So there are totally three neutrino mass sum rules, which are given by
Eqs. (\ref{27}, \ref{49}). It is found that these sum rules have no chance to be in agreement with the realistic results.

\section{One neutrino mass being vanishing}

In this section, we perform a study on the possible textures of neutrino mass matrix that can lead to
$\theta^{}_{23} = \pi/4$ and $\delta = - \pi/2$ in the scenario of one neutrino mass being vanishing. Given a vanishing neutrino mass, the neutrino mass spectrum can be fixed with the help of measured neutrino mass squared differences. On the other hand, there is only one effective Majorana CP phase which will be specified to be $\sigma$. So we just need one neutrino mass sum rule arising from the requirement of some diagonalization condition(s) holding automatically to completely determine the neutrino physical parameters.

\subsection{$m^{}_1 = 0$}

In the case of $m^{}_1 = 0$, the vanishing of ${\rm Re}\left(m^{}_1 e^{2{\rm i}\rho} \right)$ and ${\rm Im}\left(m^{}_1 e^{2{\rm i}\rho} \right)$ gives the following two new conditions
\begin{eqnarray}
&&  {\rm H:}  \quad \sin 2 \theta^{}_{12} R^{}_{\rm Y 12} = c^2_{12} R^{}_{\rm Y 11} + s^2_{12} R^{}_{\rm Y 22}  \;, \nonumber \\
&& {\rm I:}  \quad \sin 2 \theta^{}_{12} I^{}_{\rm Y 12} = c^2_{12} I^{}_{\rm Y 11} + s^2_{12} I^{}_{\rm Y 22}  \;,
\label{52}
\end{eqnarray}
in addition to those in Eq. (\ref{15}).
By relating the expressions for $\theta^{}_{12}$ derived from diagonalization conditions E and H and F and I, we obtain the following two new constraint equations in addition to those in Eqs. (\ref{16}-\ref{19})
\begin{eqnarray}
&& {\rm EH}: \quad R^{}_{\rm Y 11} + R^{}_{\rm Y 22} = {\rm sgn} \left(R^{}_{\rm Y 12}\right) \sqrt{4 \left(R^{}_{\rm Y 12}\right)^2 + \left(R^{}_{\rm Y 11} - R^{}_{\rm Y 22}\right)^2 }  \;, \nonumber \\
&& {\rm FI}: \quad I^{}_{\rm Y 11} + I^{}_{\rm Y 22} = {\rm sgn} \left(I^{}_{\rm Y 12}\right) \sqrt{4 \left(I^{}_{\rm Y 12}\right)^2 + \left(I^{}_{\rm Y 11} - I^{}_{\rm Y 22}\right)^2 } \;,
\label{53}
\end{eqnarray}
where the expressions for $R^{}_{\rm Y 12}$, $I^{}_{\rm Y 12}$, $R^{}_{\rm Y 11} - R^{}_{\rm Y 22}$ and $I^{}_{\rm Y 11} - I^{}_{\rm Y 22}$ have been given in Eq. (\ref{18}), while $R^{}_{\rm Y 11} + R^{}_{\rm Y 22}$ and $I^{}_{\rm Y 11} + I^{}_{\rm Y 22}$ are given by
\begin{eqnarray}
R^{}_{\rm Y 11} + R^{}_{\rm Y 22}  &=& \frac{\bar R^{}_{ee} - 3\bar R^{}_{\mu \tau}}{2} + \frac{\bar R^{}_{\mu \mu} + \bar R^{}_{\tau \tau}}{4} + {\rm sgn} \left(\bar I^{}_{e \mu} + \bar I^{}_{e \tau}\right) \nonumber \\
&& \times \sqrt{\frac{1}{2} \left( \bar I^{}_{e \mu} + \bar I^{}_{e \tau}\right)^2 + \frac{1}{4} \left(\bar R^{}_{ee} + \bar R^{}_{\mu \tau} + \frac{\bar R^{}_{\mu \mu} + \bar R^{}_{\tau \tau}}{2} \right)^2 }\;, \nonumber \\
I^{}_{\rm Y 11} + I^{}_{\rm Y 22}
&=&  \bar I^{}_{ee} - 2 \bar I^{}_{\mu\tau} \;.
\label{54}
\end{eqnarray}
So there are totally seven independent constraint equations.
In the following, we will study the various cases where one neutrino mass sum rule arises from the requirement of some diagonalization condition(s) holding automatically so that the only unknown physical parameter $\sigma$ can be determined. Before proceeding, we make two observations: (1) It is easy to see that diagonalization conditions E and H and F and I always simultaneously hold automatically, respectively. (2) When diagonalization conditions A(G) hold automatically, one has $I^{}_{\rm Y11} =0$, which will lead diagonalization conditions F and I to hold automatically too. So diagonalization conditions A(G), C(F) and I always simultaneously hold automatically.

In the case of diagonalization conditions A(G), C(F) and I holding automatically, there are the following six new constraint equations (i.e., those in Eqs. (\ref{20}, \ref{24})) in addition to those in Eqs. (\ref{16}-\ref{19}, \ref{53})
\begin{eqnarray}
&& \bar R^{}_{e \mu} = -\bar R^{}_{e \tau} \;,  \hspace{1cm} \bar I^{}_{e \mu} = \bar I^{}_{e \tau} \;, \hspace{1cm} \bar R^{}_{\mu \mu} = \bar R^{}_{\tau \tau} \;, \hspace{1cm} \bar I^{}_{\mu \mu} = - \bar I^{}_{\tau \tau} \;, \hspace{1cm} \bar I^{}_{ee} = \bar I^{}_{\mu \tau} = 0 \;,
\label{55}
\end{eqnarray}
which can be recombined into
\begin{eqnarray}
\bar M^{}_{e\mu} = - \bar M^*_{e\tau} \;, \hspace{1cm} \bar M^{}_{\mu\mu} = \bar M^*_{\tau\tau}  \;, \hspace{1cm}
\bar M^{}_{ee} \ {\rm and} \ \bar M^{}_{\mu\tau} \ {\rm being \ real}  \;.
\label{56}
\end{eqnarray}
As a result, only two (i.e., BD and EH) of the constraint equations in Eqs. (\ref{16}-\ref{19}, \ref{53}) are still independent ones. So there is one neutrino mass sum rule, which is directly obtained as
\begin{eqnarray}
\sigma = 0 \ {\rm or} \ \frac{\pi}{2} \;,
\label{57}
\end{eqnarray}
from Eqs. (\ref{21}, \ref{25}) by taking $m^{}_1 = 0$.
For such a result, $m^{}_\beta$ and $\Sigma$ respectively take a value of 0.009 eV and 0.059 eV.
On the other hand, $\bar M^{}_\nu$ and the magnitudes of its elements are given by
\begin{eqnarray}
\frac{\bar M^{}_{\nu}}{{\rm eV}} \simeq \begin{pmatrix}
 0.002 & 0.003+0.005 {\rm i} & -0.003+0.005 {\rm i} \\
\times & 0.028+0.001 {\rm i} & 0.022 \\
\times & \times & 0.028-0.001 {\rm i}
\end{pmatrix} \;, \hspace{1cm}
\frac{|\bar M^{}_{\nu}|}{{\rm eV}} \simeq \begin{pmatrix}
0.002 & 0.006 & 0.006 \\
\times & 0.028 & 0.022 \\
\times & \times & 0.028
\end{pmatrix}  \;,
\end{eqnarray}
for $\sigma  =0$, or
\begin{eqnarray}
\frac{\bar M^{}_{\nu}}{{\rm eV}} \simeq\begin{pmatrix}
-0.004 & -0.003+0.005 {\rm i} & 0.003+0.005 {\rm i} \\
\times & 0.022-0.001 {\rm i} & 0.028 \\
\times & \times & 0.022+0.001 {\rm i}
\end{pmatrix} \;, \hspace{1cm}
\frac{|\bar M^{}_{\nu}|}{{\rm eV}} \simeq \begin{pmatrix}
0.004 & 0.006 & 0.006 \\
\times & 0.022 & 0.028 \\
\times & \times & 0.022
\end{pmatrix}  \;,
\end{eqnarray}
for $\sigma  =\pi/2$.

In the case of diagonalization condition B holding automatically, there are two new constraint equations given by Eq. (\ref{22}) in addition to those in Eqs. (\ref{16}-\ref{19}, \ref{53}). As a result, only
six (i.e., AG, EF, EH, FI and two of AC, AD and CD) of the constraint equations in Eqs. (\ref{16}-\ref{19}, \ref{53}) are still independent ones. So there is one neutrino mass sum rule, which is directly obtained as
\begin{eqnarray}
m^{}_2 s^2_{12} \cos 2 \sigma + m^{}_3 = 0 \;,
\label{58}
\end{eqnarray}
from Eq. (\ref{23}) by taking $m^{}_1 = 0$. Apparently, this sum rule can never be fulfilled in the NO case.

In the case of diagonalization conditions D(E) and H holding automatically, there are the following four new constraint equations (i.e., those in Eq. (\ref{26}) and $R^{}_{\rm Y 11} = R^{}_{\rm Y 22} = 0$)
\begin{eqnarray}
&& \hspace{-0.7cm} \bar R^{}_{ee} -2 \bar R^{}_{\mu \tau}
= -{\rm sgn} \left(\bar I^{}_{e \mu} + \bar I^{}_{e \tau}\right)
\sqrt{2 \left( \bar I^{}_{e \mu} + \bar I^{}_{e \tau}\right)^2 + \left(\bar R^{}_{ee} + 2 \bar R^{}_{\mu \tau} \right)^2 }  \;, \nonumber \\
&& \hspace{-0.7cm} \bar I^{}_{\mu \mu} = \bar I^{}_{\tau \tau} \;, \hspace{1cm} \bar R^{}_{e \mu} =\bar R^{}_{e \tau} \;, \hspace{1cm}  \bar R^{}_{\mu\mu} + \bar R^{}_{\tau\tau} = 2 \bar R^{}_{\mu\tau}\;,
\label{59}
\end{eqnarray}
in addition to those in Eqs. (\ref{16}-\ref{19}, \ref{53}). As a result, only four (i.e., equations AG, FI and two of AB, AC and BC) of the constraint equations in Eqs. (\ref{16}-\ref{19}, \ref{53}) are still independent ones. So there is one neutrino mass sum rule, which is directly obtained as
\begin{eqnarray}
\sigma = \frac{\pi}{4} \ {\rm or} \ \frac{3\pi}{4} \;,
\label{60}
\end{eqnarray}
from Eq. (\ref{27}) by taking $m^{}_1 = 0$.
For such a result, $m^{}_\beta$ and $\Sigma$ respectively take a value of 0.009 eV and 0.059 eV.
In the case of $\sigma  =\pi/4$, $\bar M^{}_\nu$ and the magnitudes of its elements are given by
\begin{eqnarray}
\frac{\bar M^{}_{\nu}}{{\rm eV}} \simeq \begin{pmatrix}
-0.001+0.003 {\rm i} & 0.008 {\rm i} & 0.002 {\rm i} \\
\times & 0.024+0.003 {\rm i} & 0.025-0.003 {\rm i} \\
\times & \times & 0.025+0.003 {\rm i}
\end{pmatrix} \;, \hspace{0.5cm}
\frac{|\bar M^{}_{\nu}|}{{\rm eV}} \simeq \begin{pmatrix}
0.003 & 0.008 & 0.002 \\
\times & 0.024 & 0.025 \\
\times & \times & 0.025
\end{pmatrix}  \;.
\label{60-1}
\end{eqnarray}
In the case of $\sigma  =3\pi/4$, $\bar M^{}_\nu$ can be obtained by making the interchanges $\bar I^{}_{e\mu} \leftrightarrow \bar I^{}_{e\tau}$, $\bar R^{}_{\mu\mu} \leftrightarrow  \bar R^{}_{\tau\tau}$ and $\bar I^{}_{\mu\mu} \leftrightarrow  - \bar I^{}_{\tau\tau}$ and a sign change for $\bar I^{}_{ee}$ and $\bar I^{}_{\mu \tau}$ in the $\bar M^{}_\nu$ given by Eq. (\ref{60-1}).

\subsection{$m^{}_3 =0$}

In the case of $m^{}_3 =0$, the vanishing of ${\rm Im}\left(m^{}_1 e^{2{\rm i}\rho} \right)$ and ${\rm Re} (m^{}_3)$ gives the following two new conditions
\begin{eqnarray}
&& {\rm I:}  \quad \sin 2 \theta^{}_{12} I^{}_{\rm Y 12} = c^2_{12} I^{}_{\rm Y 11} + s^2_{12} I^{}_{\rm Y 22}  \;, \nonumber \\
&&  {\rm J:}  \quad \sin 2 \theta^{}_{13} I^{}_{\rm X 13} = s^2_{13} R^{}_{\rm X 11} - c^2_{13} R^{}_{\rm X 33}     \;,
\label{61}
\end{eqnarray}
in addition to those in Eq. (\ref{15}).
By relating the expressions for $\theta^{}_{13}$ and $\theta^{}_{12}$ derived from diagonalization conditions B and J and F and I, we obtain the following two new constraint equations
\begin{eqnarray}
&& \hspace{-0.3cm} {\rm BJ}: \quad \bar R^{}_{ee} - \bar R^{}_{\mu\tau} - \frac{\bar R^{}_{\mu\mu} + \bar R^{}_{\tau\tau}}{2} = {\rm sgn} \left( \bar I^{}_{e\mu} + \bar I^{}_{e\tau} \right)
\sqrt{ 2 \left( \bar I^{}_{e\mu} + \bar I^{}_{e\tau} \right)^2 + \left( \bar R^{}_{ee} + \bar R^{}_{\mu\tau} + \frac{\bar R^{}_{\mu\mu} + \bar R^{}_{\tau\tau}}{2} \right)^2 } \;, \nonumber \\
&& \hspace{-0.3cm} {\rm FI}: \quad I^{}_{\rm Y 11} + I^{}_{\rm Y 22} = {\rm sgn} \left(I^{}_{\rm Y 12}\right) \sqrt{4 \left(I^{}_{\rm Y 12}\right)^2 + \left(I^{}_{\rm Y 11} - I^{}_{\rm Y 22}\right)^2 } \;,
\label{62}
\end{eqnarray}
in addition to those in Eqs. (\ref{16}-\ref{19}). So there are totally seven independent constraint equations. In the following, we will study the various cases where one neutrino mass sum rule arises from the requirement of some diagonalization condition(s) holding automatically so that the only unknown physical parameter $\sigma$ can be determined. Before proceeding, we make two observations: (1) It is easy to see that diagonalization conditions B and J and F and I always simultaneously hold automatically, respectively. (2) When diagonalization conditions A(G) hold automatically, one has $I^{}_{\rm Y11} =0$, which will lead diagonalization conditions F and I to hold automatically too. So diagonalization conditions A(G), C(F) and I always simultaneously hold automatically.

In the case of diagonalization conditions A(G), C(F) and I holding automatically, there are six new constraint equations given by Eq. (\ref{55})
in addition to those in Eqs. (\ref{16}-\ref{19}, \ref{62}).
As a result, only two (i.e., BD and BJ) of the constraint equations in Eqs. (\ref{16}-\ref{19}, \ref{62}) are still independent ones. So there is one neutrino mass sum rule, which is directly obtained from Eqs. (\ref{21}, \ref{25}) by taking $\rho = 0$ and is the same as that in Eq. (\ref{57}).
For such a result, $m^{}_\beta$ and $\Sigma$ respectively take a value of 0.049 eV and 0.10 eV.
On the other hand, $\bar M^{}_\nu$ and the magnitudes of its elements are given by
\begin{eqnarray}
\frac{\bar M^{}_{\nu}}{{\rm eV}} \simeq \begin{pmatrix}
0.049 & 0.005 {\rm i} & 0.005 {\rm i} \\
\times & 0.024 & -0.025 \\
\times & \times & 0.024
\end{pmatrix} \;, \hspace{1cm}
\frac{|\bar M^{}_{\nu}|}{{\rm eV}} \simeq \begin{pmatrix}
0.049 & 0.005  & 0.005  \\
\times & 0.024 & 0.025 \\
\times & \times & 0.024
\end{pmatrix}  \;,
\end{eqnarray}
for $\sigma  =0$, or
\begin{eqnarray}
\frac{\bar M^{}_{\nu}}{{\rm eV}} \simeq\begin{pmatrix}
0.019 & -0.032+0.002 {\rm i} & 0.032+0.002 {\rm i} \\
\times & -0.010-0.007 {\rm i} & 0.010 \\
\times & \times & -0.010+0.007 {\rm i}
\end{pmatrix} \;, \hspace{1cm}
\frac{|\bar M^{}_{\nu}|}{{\rm eV}} \simeq \begin{pmatrix}
0.019 & 0.032 & 0.032 \\
\times & 0.012 & 0.010 \\
\times & \times & 0.012
\end{pmatrix}  \;,
\end{eqnarray}
for $\sigma  =\pi/2$.

In the case of diagonalization conditions B and J holding automatically, there are the following three new constraint equations (i.e., those in Eq. (\ref{22}) and $R^{}_{\rm X 11} = R^{}_{\rm X 33} = 0$)
\begin{eqnarray}
\bar I^{}_{e \mu} = - \bar I^{}_{e \tau} \;, \hspace{1cm} -2 \bar R^{}_{\mu\tau}  =  \bar R^{}_{\mu \mu} + \bar R^{}_{\tau \tau} \;, \hspace{1cm} \bar R^{}_{ee} =  0 \;,
\label{63}
\end{eqnarray}
in addition to those in Eqs. (\ref{16}-\ref{19}, \ref{62}).
As a result, only five (i.e., AG, EF, FI and two of AC, AD and CD) of the constraint equations in Eqs. (\ref{16}-\ref{19}, \ref{62}) are still independent ones. So there is one neutrino mass sum rule, which is directly obtained as
\begin{eqnarray}
m^{}_1 c^2_{12} + m^{}_2 s^2_{12} \cos 2 \sigma = 0 \;,
\label{64}
\end{eqnarray}
from Eq. (\ref{23}) by taking $\rho = m^{}_3 = 0$. Apparently, this sum rule can never be fulfilled in the IO case.

In the case of diagonalization conditions D(E) holding automatically, there are three new constraint equations given by Eq. (\ref{26})
in addition to those in Eqs. (\ref{16}-\ref{19}, \ref{62}).
As a result, only five (i.e., AG, BJ, FI and two of AB, AC and BC) of the constraint equations in Eqs. (\ref{16}-\ref{19}, \ref{62}) are still independent ones. So there is one neutrino mass sum rule, which is directly obtained as
\begin{eqnarray}
m^{}_1 - m^{}_2 \cos 2 \sigma = 0 \;,
\label{65}
\end{eqnarray}
from Eq. (\ref{27}) by taking $\rho = m^{}_3 = 0$.
By solving this equation, one obtains $\sigma = 0.03 \pi$ or $0.97 \pi$.
For such a result, $m^{}_\beta$ and $\Sigma$ respectively take a value of 0.049 eV and 0.10 eV.
In the case of $\sigma \simeq 0.03 \pi$, $\bar M^{}_\nu$ and the magnitudes of its elements are given by
\begin{eqnarray}
\frac{\bar M^{}_{\nu}}{{\rm eV}} \simeq \begin{pmatrix}
0.048+0.003 {\rm i} & 0.008 {\rm i} & 0.002 {\rm i} \\
\times & 0.024+0.003 {\rm i} & -0.025-0.003 {\rm i} \\
\times & \times & 0.025+0.003 {\rm i}
\end{pmatrix} \;, \hspace{0.5cm}
\frac{|\bar M^{}_{\nu}|}{{\rm eV}} \simeq \begin{pmatrix}
0.048 & 0.008 & 0.002 \\
\times & 0.024 & 0.025 \\
\times & \times & 0.025
\end{pmatrix}  \;.
\label{65-1}
\end{eqnarray}
In the case of $\sigma  \simeq 0.97 \pi$, $\bar M^{}_\nu$ can be obtained by making the interchanges $\bar I^{}_{e\mu} \leftrightarrow \bar I^{}_{e\tau}$, $\bar R^{}_{\mu\mu} \leftrightarrow  \bar R^{}_{\tau\tau}$ and $\bar I^{}_{\mu\mu} \leftrightarrow  - \bar I^{}_{\tau\tau}$ and a sign change for $\bar I^{}_{ee}$ and $\bar I^{}_{\mu \tau}$ in the $\bar M^{}_\nu$ given by Eq. (\ref{65-1}).

\section{Discussions}

In this section, we give some discussions about the possible textures of neutrino mass matrix that can lead to $\theta^{}_{23} = \pi/4$, $\delta = - \pi/2$ and maximal Majorana CP phases as well as the model realization and breakings of the obtained textures.

\subsection{$\rho, \sigma = \pi/4$ or $3\pi/4$}

Motivated by the $\mu$-$\tau$ reflection symmetry which predicts $\theta^{}_{23} = \pi/4$, $\delta = - \pi/2$ and trivial Majorana CP phases, we make an attempt to derive the possible textures of neutrino mass matrix that can lead to $\theta^{}_{23} = \pi/4$, $\delta = - \pi/2$ and maximal Majorana CP phases. In this scenario, the vanishing of ${\rm Re}\left(m^{}_1 e^{2{\rm i}\rho} \right)$ and ${\rm Re}\left(m^{}_2 e^{2{\rm i}\sigma} \right)$ gives the following two new conditions
\begin{eqnarray}
&&  {\rm H:}  \quad \sin 2 \theta^{}_{12} R^{}_{\rm Y 12} = c^2_{12} R^{}_{\rm Y 11} + s^2_{12} R^{}_{\rm Y 22}  \;, \nonumber \\
&& {\rm K:}  \quad \sin 2 \theta^{}_{12} R^{}_{\rm Y 12} = -s^2_{12} R^{}_{\rm Y 11} - c^2_{12} R^{}_{\rm Y 22}  \;,
\label{66}
\end{eqnarray}
in addition to those in Eq. (\ref{15}).
A combination of diagonalization conditions E, H and K results in $R^{}_{\rm Y 11} = R^{}_{\rm Y 22} = R^{}_{\rm Y 12} =0$, which lead diagonalization conditions D(E), H and K to hold automatically. Hence there are four new constraint equations given by Eq. (\ref{59}) in addition to those in Eqs. (\ref{16}-\ref{19}). As a result, only three (i.e., AG and two of AB, AC and BC) of the constraint equations in Eqs. (\ref{16}-\ref{19}) are still independent ones. So there are totally seven independent constraint equations.
In the following, we will study the various cases where one neutrino mass sum rule arises from the requirement of some diagonalization condition(s) holding automatically so that the only unknown physical parameter (the absolute neutrino mass scale) can be determined.

In the case of diagonalization conditions A(G) (in addition to D(E), H and K) holding automatically, there are the following seven new constraint equations (i.e., those in Eqs. (\ref{20}, \ref{59}))
\begin{eqnarray}
&& \hspace{-0.7cm} \bar R^{}_{ee} -2 \bar R^{}_{\mu \tau}
= -{\rm sgn} \left(\bar I^{}_{e \mu} + \bar I^{}_{e \tau}\right)
\sqrt{2 \left( \bar I^{}_{e \mu} + \bar I^{}_{e \tau}\right)^2 + \left(\bar R^{}_{ee} + 2 \bar R^{}_{\mu \tau} \right)^2 }  \;, \nonumber \\
&& \hspace{-0.7cm} \bar I^{}_{\mu \mu} = \bar I^{}_{\tau \tau} = - \bar I^{}_{\mu \tau} \;, \hspace{1cm} \bar R^{}_{e \mu} =\bar R^{}_{e \tau} = \bar I^{}_{ee} = 0 \;, \hspace{1cm}  \bar R^{}_{\mu\mu} + \bar R^{}_{\tau\tau} = 2 \bar R^{}_{\mu\tau}\;,
\label{67}
\end{eqnarray}
in addition to those in Eqs. (\ref{16}-\ref{19}). As a result, only one (i.e., BC) of the constraint equations in Eqs. (\ref{16}-\ref{19}) is still an independent one. So there is one neutrino mass sum rule, which is directly obtained as
\begin{eqnarray}
m^{}_1 c^2_{12} \pm m^{}_2 s^2_{12} = 0 \;,
\label{68}
\end{eqnarray}
from Eq. (\ref{21}) by taking $\rho, \sigma = \pi/4$ or $3\pi/4$. By solving this equation, one obtains $m^{}_1=0.004$ eV with $[\rho, \sigma] = [\pi/4, 3\pi/4]$ or $[3\pi/4, \pi/4]$.
For such a result, $m^{}_\beta$ and $\Sigma$ respectively take a value of 0.010 eV and 0.064 eV.
In the case of $[\rho, \sigma] = [\pi/4, 3\pi/4]$, $\bar M^{}_\nu$ and the magnitudes of its elements are given by
\begin{eqnarray}
\frac{\bar M^{}_{\nu}}{{\rm eV}} \simeq \begin{pmatrix}
 -0.001 & 0.001 {\rm i} & 0.010 {\rm i} \\
\times & 0.026-0.003 {\rm i} & 0.025+0.003 {\rm i} \\
\times & \times & 0.024-0.003 {\rm i}
\end{pmatrix} \;, \hspace{0.5cm}
\frac{|\bar M^{}_{\nu}|}{{\rm eV}} \simeq \begin{pmatrix}
0.001 & 0.001  & 0.010  \\
\times & 0.026 & 0.025 \\
\times & \times & 0.024
\end{pmatrix}  \;.
\label{68-1}
\end{eqnarray}
In the case of $[\rho, \sigma] = [3\pi/4, \pi/4]$, $\bar M^{}_\nu$ can be obtained by making the interchanges $\bar I^{}_{e\mu} \leftrightarrow \bar I^{}_{e\tau}$, $\bar R^{}_{\mu\mu} \leftrightarrow  \bar R^{}_{\tau\tau}$ and $\bar I^{}_{\mu\mu} \leftrightarrow  - \bar I^{}_{\tau\tau}$ and a sign change for $\bar I^{}_{\mu \tau}$ in the $\bar M^{}_\nu$ given by Eq. (\ref{68-1}).

In the case of diagonalization condition B (in addition to D(E), H and K) holding automatically, there are six new constraint equations given by Eqs. (\ref{22}, \ref{59})
in addition to those in Eqs. (\ref{16}-\ref{19}). As a result, only two (i.e., AC and AG) of the constraint equations in Eqs. (\ref{16}-\ref{19}) are still independent ones. So there is one neutrino mass sum rule, which is directly obtained as
\begin{eqnarray}
 m^{}_3 = 0 \;,
\label{69}
\end{eqnarray}
from Eq. (\ref{23}) by taking $\rho, \sigma = \pi/4$ or $3\pi/4$. However, as discussed in section 4, $\rho$ would be fixed to 0 in the case of $m^{}_3 =0$. So this sum rule has no chance to be in agreement with the realistic results.

In the case of diagonalization conditions C(F) (in addition to D(E), H and K) holding automatically,
there are six new constraint equations given by Eqs. (\ref{24}, \ref{59})
in addition to those in Eqs. (\ref{16}-\ref{19}). As a result, only two (i.e., AB and AG) of the constraint equations in Eqs. (\ref{16}-\ref{19}) are still independent ones. So there is one neutrino mass sum rule, which is directly obtained as
\begin{eqnarray}
m^{}_1 \pm m^{}_2 = 0 \;,
\label{70}
\end{eqnarray}
from Eq. (\ref{25}) by taking $\rho, \sigma = \pi/4$ or $3\pi/4$.
Apparently, this sum rule has no chance to be in agreement with the realistic results.

\subsection{Model realization}

The approach adopted by us is a bottom-up one in the sense that we start from the experimental hints for $\theta^{}_{23} \simeq \pi/4$ and $\delta \simeq -\pi/2$ and then derive the possible textures of neutrino mass matrix that can lead to maximal $\theta^{}_{23}$ and $\delta$. (In a top-down approach one starts from a specific flavor theory and then derives its phenomenological consequences.) Although all the obtained textures are on an equal footing in our approach, a particular texture will gain a more solid foundation if it can find an origin from some flavor symmetry. From the model realization point of view, the obtained textures can be classified into the following three categories.

First of all, we note that some of the obtained textures can find a connection with the $\mu$-$\tau$ reflection symmetry. One can see that the resulting neutrino mass sum rules in the case of diagonalization conditions A(G) and C(F) holding automatically in combination with $\bar M^{}_{ee} =0$ ($\bar M^{}_{\mu\tau} =0$) are the same as those in the case of $\mu$-$\tau$ reflection symmetry with $M^{}_{ee} =0$ ($M^{}_{\mu\tau} =0$) \cite{zero}. In fact, if one restores the unphysical phases with the help of $\bar M^{}_{\alpha \beta} = M^{}_{\alpha \beta} e^{-{\rm i}(\phi^{}_\alpha+ \phi^{}_\beta)}$ and takes $\phi^{}_e = \pi/2$ and $\phi^{}_\mu =  -\phi^{}_\tau$, the texture obtained in the former case will reproduce that in the latter case. For example, under such a specification for the unphysical phases, the texture of $\bar M^{}_\nu$ in Eq. (\ref{39}) gives the following texture of $M^{}_\nu$
\begin{eqnarray}
M^{}_{e\mu} = M^*_{e\tau} \;, \hspace{1cm} M^{}_{\mu\mu} = M^*_{\tau\tau}  \;, \hspace{1cm}
M^{}_{ee} =0 \;, \hspace{1cm} M^{}_{\mu\tau} \ {\rm being \ real}  \;.
\label{70.1}
\end{eqnarray}
(BD just gives an expression for the unphysical phase $\phi^{}_\mu$ in terms of the neutrino mass matrix elements.) As a comparison, if we take $\phi^{}_e = 0$ and $\phi^{}_\mu =  -\phi^{}_\tau$, Eq. (\ref{39}) gives
\begin{eqnarray}
M^{}_{e\mu} = - M^*_{e\tau} \;, \hspace{1cm} M^{}_{\mu\mu} = M^*_{\tau\tau}  \;, \hspace{1cm}
M^{}_{ee} =0 \;, \hspace{1cm} M^{}_{\mu\tau} \ {\rm being \ real}  \;.
\label{70.2}
\end{eqnarray}
These two textures give the same results for the physical parameters. But the former one has the advantage of having a connection with the $\mu$-$\tau$ reflection symmetry. Of course, one can choose any other specification for the unphysical phases, which will not alter the results for the physical parameters but may lead to a different texture.
Similarly, the texture obtained in the case of diagonalization conditions A(G) and C(F) holding automatically in combination with $m^{}_1$ or $m^{}_3=0$ can reproduce that from the $\mu$-$\tau$ reflection symmetry embedded in the minimal seesaw \cite{mtinms}.

A texture belonging to the second category is one that can not be connected to a known flavor symmetry like the $\mu$-$\tau$ reflection symmetry but only possesses linear relations among the neutrino mass matrix elements. It may be realized with the help of some flavor symmetry. An example is the texture obtained in the case of diagonalization conditions A(G) and B holding automatically in combination with $\bar M^{}_{ee} =0$. If one restores the unphysical phases and takes $\phi^{}_\mu =  \phi^{}_\tau + \pi$, Eq. (\ref{37}) gives
\begin{eqnarray}
&& M^{}_{e \mu} = M^{}_{e \tau} \;,  \hspace{1cm} 2 M^{}_{\mu \tau} =  M^{}_{\mu \mu} + M^{}_{\tau \tau} \;, \hspace{1cm} M^{}_{ee} =0 \;.
\label{71}
\end{eqnarray}
(CD and EF just give the expressions for the unphysical phases $\phi^{}_e$ and $\phi^{}_\mu $ in terms of the neutrino mass matrix elements.) The texture zero $M^{}_{ee} =0$ can be attributed to an Abelian flavor symmetry \cite{abelian} while the linear relations $M^{}_{e \mu} = M^{}_{e \tau}$ and $2 M^{}_{\mu \tau} =  M^{}_{\mu \mu} + M^{}_{\tau \tau}$ may find an origin from some non-Abelian flavor symmetry \cite{review}.

A texture belonging to the third category is one that possesses not only linear but also non-linear relations among the neutrino mass matrix elements. It can only be partially realized with the help of some flavor symmetry. Let us take the texture obtained in the case of diagonalization conditions A(G) and D(E) holding automatically in combination with $\bar M^{}_{ee} =0$ as an example. If one restores the unphysical phases and takes $\phi^{}_e =  0$ and $\phi^{}_\mu =  - \phi^{}_\tau = \pi/2$, Eq. (\ref{40}) gives
\begin{eqnarray}
&& R^{}_{\mu \tau} + \frac{3}{2} \left(R^{}_{\mu \mu} + R^{}_{\tau \tau}\right)
= {\rm sgn} \left( R^{}_{e \mu} - R^{}_{e \tau}\right)
\sqrt{2 \left( R^{}_{e \mu} - R^{}_{e \tau}\right)^2 + \left( R^{}_{\mu \tau} - \frac{ R^{}_{\mu \mu} + R^{}_{\tau \tau} } {2}\right)^2 }  \;, \nonumber \\
&&  R^{}_{ee} = I^{}_{e \mu} = I^{}_{e \tau} = I^{}_{ee} = 0 \;, \hspace{1cm}   I^{}_{\mu \mu}  =  I^{}_{\tau \tau} =  I^{}_{\mu \tau} \;,
\label{72}
\end{eqnarray}
with $R^{}_{\alpha \beta} = {\rm Re}( M^{}_{\alpha \beta})$ and $I^{}_{\alpha \beta} = {\rm Im}( M^{}_{\alpha \beta} )$, while BC gives
\begin{eqnarray}
\left(R^{}_{e \mu} + R^{}_{e \tau}\right) \left(R^{}_{\mu \mu} - R^{}_{\tau \tau}\right)
\left(R^{}_{\mu \tau} - \frac{R^{}_{\mu \mu} + R^{}_{\tau \tau}}{2} \right)
= & & \left(R^{}_{e \mu} - R^{}_{e \tau}\right) \Big[ \left(R^{}_{e \mu} + R^{}_{e \tau}\right)^2 \nonumber \\
& & - \frac{1}{2} \left( R^{}_{\mu \mu} - R^{}_{\tau \tau}\right)^2 \Big] \;.
\label{73}
\end{eqnarray}
In this case, the imaginary part of $M^{}_\nu$ has a simplest structure which may be easily realized with the help of some flavor symmetry. However, to our knowledge, flavor symmetries are unable to give non-linear relations like those two for the real part of $M^{}_\nu$ (i.e., the first one in Eq. (\ref{72}) and that in Eq. (\ref{73})).

\subsection{Breakings}

In order to accommodate the deviations of $\theta^{}_{23}$ and $\delta$ from $\pi/4$ and $-\pi/2$, one has to consider the breakings of the obtained textures.
In the literature, at least four scenarios for the breakings of flavor symmetries (neutrino mass matrix textures) have been considered: (1) The flavor symmetries are usually introduced at an extremely high energy scale (e.g., the seesaw scale). So the renormalization group (RG) running effect should be taken into account when one confronts the flavor symmetry models with the low energy data \cite{rge}, which may induce the breakings of flavor symmetries. For example, in the RG running process the significant hierarchy between the mass of muon and that of tau can give rise to the breaking of $\mu$-$\tau$ symmetry \cite{rge2}. (2) As in this work, most of the studies on the textures of neutrino mass matrix have been performed in the basis of charged lepton mass matrix $M^{}_l$ being diagonal. But in a realistic model this may not be the case. (For example, in a grand unified theory inspired model, $M^{}_l$ will be associated with the down-type quark mass matrix which is generally treated as a non-diagonal one \cite{gut}.) In this situation, even if the special texture of $M^{}_\nu$ is realized exactly, the neutrino mixing will receive corrections from the charged lepton sector \cite{ml}. (3) The LSND experiment \cite{lsnd} and reactor antineutrino anomaly \cite{reactor} indicate the existence of eV scale sterile neutrinos mixing with the three active neutrinos. If this turns out to be true, the sterile neutrino sector may provide a source for the breakings of flavor symmetries in the active neutrino sector \cite{sterile}. (4) In concrete flavor symmetry models, it is common that the special texture of $M^{}_\nu$ only holds at the leading order but breaks to some extent at the higher orders \cite{review}.

A comprehensive study about the breakings of all the obtained textures in the above scenarios is model-dependent and beyond the scope of this paper. As an example, we give a model-independent phenomenological study of the breakings of the texture given by the $\mu$-$\tau$ reflection symmetry with $M^{}_{ee} =0$ ($M^{}_{\mu\tau} =0$). (A discussion about the breakings of the texture given by the $\mu$-$\tau$ reflection symmetry embedded in the minimal seesaw can be found in Ref. \cite{mtinms}.)
Corresponding to the four symmetry conditions in Eq. (\ref{8}) one by one, the following four dimensionless parameters can be introduced to measure the breaking strengths of $\mu$-$\tau$ reflection symmetry \cite{MTRb}
\begin{eqnarray}
\epsilon^{}_{1}=\displaystyle \frac{M^{}_{e \mu}- M^{*}_{e\tau}}{M^{}_{e \mu}+M^{*}_{e\tau}} \;, \hspace{1cm}
\epsilon^{}_{2}=\displaystyle \frac{M^{}_{\mu\mu}-M^{*}_{\tau \tau}}{M^{}_{\mu \mu}+ M^{*}_{\tau\tau}} \;, \hspace{1cm}
\epsilon^{}_{3}=\displaystyle \frac{I^{}_{ee}}{R^{}_{ee}} \;, \hspace{1cm}
\epsilon^{}_{4}=\displaystyle \frac{I^{}_{\mu \tau}}{R^{}_{\mu \tau}} \;,
\label{74}
\end{eqnarray}
whose magnitudes should be small enough (e.g., $\le 0.1$) in order to keep the symmetry as an approximate one. Two immediate comments are given as follows: (1) A model-specific breaking is characterized by a given pattern of these four symmetry-breaking parameters. For example, a RG induced breaking of $\mu$-$\tau$ reflection symmetry is characterized by
$I^{}_{1,2} = \epsilon^{}_{3,4}=0$ (for $I^{}_{1,2} = {\rm Im}( \epsilon^{}_{1,2})$) and $R^{}_{2} = 2 R^{}_{1}$ (for $R^{}_{1,2} = {\rm Re}( \epsilon^{}_{1,2})$) \cite{MTRb}. Correspondingly, the implications of a model-specific breaking can be inferred from the general model-independent results.
(2) It has been noted that $\epsilon^{}_3$ and $\epsilon^{}_4$ are respectively equivalent to an $I^{}_{1} \simeq -\epsilon^{}_3/2$ and an $I^{}_{1} \simeq -\epsilon^{}_4/2$ plus an $I^{}_{2} \simeq -\epsilon^{}_4$
to a good approximation,
thereby allowing one to pay attention to $\epsilon^{}_{1,2}$ \cite{MTRb}.
But for the condition of $R^{}_{ee} =0$ ($R^{}_{\mu\tau} =0$), there is not a well-defined dimensionless parameter that can be used to measure its breaking strength, so we will preserve it.
In the following, given a small value of $R^{}_{1,2}$ or $I^{}_{1,2}$, we will study the deviations of physical parameters
\begin{eqnarray}
&& \Delta m^{}_{1,3} = m^{}_{1,3} - m^{(0)}_{1,3} \;, \hspace{1cm}
\Delta \theta^{}_{23} = \theta^{}_{23} - \theta^{(0)}_{23} \;, \hspace{1cm}
\Delta \delta  = \delta^{} - \delta^{(0)} \;, \nonumber \\
&& \Delta \rho  = \rho^{} - \rho^{(0)} \;, \hspace{1cm}
\Delta \sigma  = \sigma^{} - \sigma^{(0)} \;,
\label{75}
\end{eqnarray}
from their values in the symmetry limit (which are labelled by a superscript ``(0)").

In the case of the $\mu$-$\tau$ reflection symmetry with $M^{}_{ee} =0$, the deviations of physical parameters induced by a $R^{}_{1,2}$ or $I^{}_{1,2} $ of the benchmark value 0.1 are given in Table 1.
The results outside (inside) the brackets are obtained in the case of $m^{(0)}_1 = 0.006$ eV with $[\rho^{(0)}, \sigma^{(0)}] = [0, \pi/2]$ ($m^{(0)}_1 = 0.002$ eV with $[\rho^{(0)}, \sigma^{(0)}] = [\pi/2, 0]$).
It is found that $\Delta m^{}_1/m^{(0)}_{1}$ is quite small, which can be understood in a way as follows: The $\mu$-$\tau$ reflection symmetry itself is unable to give a prediction for the neutrino masses. It is $M^{}_{ee} =0$ that helps us fix the neutrino mass spectrum. So the preservation of $M^{}_{ee} =0$ ensures a small $\Delta m^{}_1/m^{(0)}_{1}$. On the other hand, a considerable $\Delta \theta^{}_{23}$ (significant $\Delta \delta$) may arise from $R^{}_2$ ($R^{}_1$ and $I^{}_2$). Because of having distinct main origins, a considerable $\Delta \theta^{}_{23}$ does not necessarily signify a significant $\Delta \delta$, and vice versa. In magnitude, given small values of $R^{}_{1,2}$ and $I^{}_{1,2} $, $\Delta \theta^{}_{23}$ is not more than a few degrees while $\Delta \delta$ can reach dozens of degrees. Furthermore, all the deviations of CP phases induced by $I^{}_2$ can be significant, while all the deviations of physical parameters induced by $I^{}_1$ are negligibly small.

\begin{table}[h]
\centering
\begin{tabular}{|p{1.5cm}<{\centering}|p{2.5cm}<{\centering}|p{2.1cm}<{\centering}|p{2.5cm}<{\centering}
|p{2.1cm}<{\centering}|p{2.1cm}<{\centering}|} \hline
& $\Delta m^{}_{1}$/$m^{(0)}_1$ & $\Delta \theta^{}_{23}$ & $\Delta \delta$ &  $\Delta \rho$  & $\Delta \sigma$ \\ \hline
$R^{}_1 =0.1$ & $-0.04$ (0.07) & 0.00 (0.00) & $-0.29$ (0.14) & 0.09 (0.11) & 0.02 (0.01)  \\ \hline
$I^{}_1 =0.1$ & 0.00 (0.00)  & 0.00 (0.00) & 0.00 (0.00) & 0.00 (0.00) & 0.00 (0.00) \\ \hline
$R^{}_2 =0.1$ & 0.00 (0.00) & 0.04 (0.06) & $-0.01$ ($-0.03$) & 0.01 ($-0.03$) & 0.01 ($-0.01$) \\ \hline
$I^{}_2 =0.1$ & $-0.05$ (0.08) & 0.00 (0.00) & 0.57 ($-0.38$) & $-0.33$ (0.11) & $-0.26$ (0.21) \\ \hline
\end{tabular}
\caption{In the case of the $\mu$-$\tau$ reflection symmetry with $M^{}_{ee} =0$, the deviations of physical parameters induced by a $R^{}_{1,2}$ or $I^{}_{1,2} $ of the benchmark value 0.1.}
\end{table}

In the case of the $\mu$-$\tau$ reflection symmetry with $M^{}_{\mu\tau} =0$, the deviations of physical parameters induced by a $R^{}_{1,2}$ or $I^{}_{1,2} $ of some benchmark values are given in Table 2.
The results outside (inside) the brackets are obtained in the case of $m^{(0)}_1 = 0.165$ eV with $[\rho^{(0)}, \sigma^{(0)}] = [0, 0]$ ($m^{(0)}_3 = 0.021$ eV with $[\rho^{(0)}, \sigma^{(0)}] = [\pi/2, 0]$). In most cases, even a tiny $R^{}_{1,2}$ or $I^{}_{1,2}$ (for which we specify 0.01 or 0.005 as the benchmark value) can induce rather large deviations of physical parameters.
For a similar reason (i.e., the preservation of $M^{}_{\mu\tau} =0$), $\Delta m^{}_{1,3}/m^{(0)}_{1,3}$ are quite small. In the case of $m^{(0)}_1 = 0.165$ eV with $[\rho^{(0)}, \sigma^{(0)}] = [0, 0]$, $R^{}_{1,2}$ can induce significant $\Delta \theta^{}_{23}$ but much smaller $\Delta \delta$, $\Delta \rho$ and $\Delta \sigma$, while $I^{}_{1,2}$ can induce significant $\Delta \delta$ but negligibly small deviations of other physical parameters.
In the case of $m^{(0)}_3 = 0.021$ eV with $[\rho^{(0)}, \sigma^{(0)}] = [\pi/2, 0]$, all of $R^{}_{1,2}$ and $I^{}_{1,2}$ can induce significant deviations of CP phases. Unfortunately, such a case can not accommodate a considerable $\Delta \theta^{}_{23}$: One could obtain a sizable $\Delta \theta^{}_{23}$ by increasing the value of $R^{}_1$ from the benchmark value 0.01. But at the same time $\delta$, $\rho$ and $\sigma$ would go far away from their values in the symmetry limit.

\begin{table}[h]
\centering
\begin{tabular}{|p{2.7cm}<{\centering}|p{2.1cm}<{\centering}|p{2.1cm}<{\centering}|p{2.5cm}<{\centering}
|p{2.1cm}<{\centering}|p{2.1cm}<{\centering}|} \hline
& $\Delta m^{}_{1,3}$/$m^{(0)}_{1,3}$ & $\Delta \theta^{}_{23}$ & $\Delta \delta$ &  $\Delta \rho$  & $\Delta \sigma$ \\ \hline
$R^{}_1 =0.1$ (0.01) & 0.00 (0.02) & 0.10 ($-0.01$) & $0.02$ ($-0.21$) & 0.00 (0.15) & 0.00 (0.07)  \\ \hline
$I^{}_1 =0.005$ & 0.02 (0.05)   & 0.00 (0.00)  & $-0.22$ (0.31) & 0.00 ($-0.22$)  & 0.00 ($-0.09$) \\ \hline
$R^{}_2 =0.01$ & $-0.06$ (0.01) & 0.20 (0.00) & $-0.01$ ($-0.18$) & 0.05 ($0.10$) & $-0.02$ ($0.05$) \\ \hline
$I^{}_2 =0.01$ & $0.02$ (0.05) & 0.00 (0.00) & 0.22 ($-0.31$) & 0.00 (0.22) & 0.00 (0.09) \\ \hline
\end{tabular}
\caption{In the case of the $\mu$-$\tau$ reflection symmetry with $M^{}_{\mu\tau} =0$, the deviations of physical parameters induced by a $R^{}_{1,2}$ or $I^{}_{1,2} $ of some benchmark values.}
\end{table}

Finally, we point out that the above results can be understood in an analytical approximation way by expanding the parameters around their values in the symmetry limit \cite{MTRb}. At the first order of parameter deviations, $\Delta m^{}_{1,3}$ are vanishing while the deviations of other parameters are linear functions of $R^{}_{1,2}$ and $I^{}_{1,2}$. Only up to the second order of parameter deviations can $\Delta m^{}_{1,3}$ receive some non-vanishing contributions, making its dependence on $R^{}_{1,2}$ and $I^{}_{1,2}$ be of a quadratic form \footnote{This explains why $m^{}_{1,3}$ are stable against the symmetry breakings.}.
Provided that the parameter deviations remain small enough, the results for another value of $R^{}_{1,2}$ or $I^{}_{1,2}$ can be approximately obtained from those in Tables 1-2 by invoking the quadratic (linear) dependence of $\Delta m^{}_{1,3}$ (other parameters) on $R^{}_{1,2}$ and $I^{}_{1,2}$ \cite{MTRb}. When a given $R^{}_{1,2}$ or $I^{}_{1,2}$ changes its sign, $\Delta m^{}_{1,3}$ remain invariant while the deviations of other parameters also undergo a sign change.

\section{Summary}

\begin{table}[h]
\centering
\begin{tabular}{|p{2.5cm}<{\centering}|p{3.5cm}<{\centering}|p{7.5cm}<{\centering}|} \hline
\multirow{4}{*}{$\bar M^{}_{ee} = 0$}
 & A(G), B & $\times$  \\ \cline{2-3}
 & \multirow{2}{*}{A(G) , C(F)} & $m^{}_1 = 0.006$ eV, $\rho = 0$, $\sigma =\pi/2$   \\ \cline{3-3}  && $m^{}_1 = 0.002$ eV, $\rho = \pi/2$, $\sigma =0$ \\ \cline{2-3}
 & A(G) , D(E) & $m^{}_1 = 0.004$ eV, $\rho = 0.79\pi$, $\sigma =0.23\pi$ \\ \hline
\multirow{3}{*}{$\bar M^{}_{e\mu} = 0$}
& \multirow{2}{*}{A(G) , D(E)} &  $m^{}_1 = 0.009$ eV, $\rho = 0.41\pi$, $\sigma =0.65\pi$ \\ \cline{3-3} && $m^{}_3 = 0.006$ eV, $\rho = 0.51\pi$, $\sigma =0.47\pi$ \\ \cline{2-3}
 & B , C(F) & $m^{}_3 = 0.0007$ eV, $\rho = 0.27\pi$, $\sigma =0.22\pi$ \\ \hline
\multirow{4}{*}{$\bar M^{}_{\mu\mu} = 0$}
& A(G) & $m^{}_3 = 0.024$ eV, $\rho = 0.97\pi$, $\sigma =0.43\pi$  \\ \cline{2-3}
& B & $m^{}_3 = 0.050$ eV, $\rho = 0.35\pi$, $\sigma =0.54\pi$  \\ \cline{2-3}
& C(F) & $m^{}_3 = 0.022$ eV, $\rho = 0.04\pi$, $\sigma =0.46\pi$  \\ \cline{2-3}
& D(E) & $\times$  \\ \hline
\multirow{3}{*}{$\bar M^{}_{\mu\tau} = 0$}
& \multirow{2}{*}{A(G) , C(F)} & $m^{}_1 = 0.165$ eV, $\rho = 0$, $\sigma =0$  \\ \cline{3-3}
&& $m^{}_3 = 0.021$ eV, $\rho = \pi/2$, $\sigma =0$ \\ \cline{2-3}
& B & $\times$  \\ \cline{2-3}
& D(E) & $\times$  \\ \hline
\multirow{3}{*}{$m^{}_1 = 0$}
& A(G) , C(F) & $\sigma =0$ or $\pi/2$  \\ \cline{2-3}
& B & $\times$  \\ \cline{2-3}
& D(E) & $\sigma = \pi/4$ or $3\pi/4$  \\ \hline
\multirow{3}{*}{$m^{}_3 = 0$}
& A(G) , C(F) & $\sigma =0$ or $\pi/2$  \\ \cline{2-3}
& B & $\times$  \\ \cline{2-3}
& D(E) & $\sigma = 0.03\pi$ or $0.97\pi$  \\ \hline
\multirow{3}{*}{maximal $\rho$/$\sigma$}
& A(G) , D(E) & $m^{}_1 = 0.004$ eV  \\ \cline{2-3}
& B , D(E) & $\times$  \\ \cline{2-3}
& C(F) , D(E) & $\times$  \\ \hline
\end{tabular}
\caption{A summary of the results for the unknown physical parameters in the various cases. In the second column, the phrase ``A(G), B'' (and so on) is used to stand for the cases of diagonalization conditions A(G) and B (and so on) holding automatically.}
\end{table}

To summarize, the purpose of this work is to derive the possible textures of neutrino mass matrix that can lead to $\theta^{}_{23} =\pi/4$ and $\delta  = -\pi/2$ in two phenomenologically appealing scenarios: (1) one neutrino mass matrix element being vanishing (2) one neutrino mass being vanishing.
In the former scenario, there are two neutrino mass sum rules arising, which can be directly read from Eq. (\ref{32}). In the latter scenario, one has $m^{}_1= \rho =0$ in the NO case or $m^{}_3 = \rho =0$ in the IO case, leaving us with only one unknown physical parameter $\sigma$ to determine. Furthermore, we also make an attempt to derive the possible textures of neutrino mass matrix that can lead to $\theta^{}_{23} =\pi/4$, $\delta  = -\pi/2$ and maximal Majorana CP phases, in which scenario there is only the absolute neutrino mass scale to be determined.
In these three scenarios, we study the various cases where one more neutrino mass sum rule arises from the requirement of some diagonalization condition(s) holding automatically so that all the three unknown physical parameters (i.e., the absolute neutrino mass scale and two Majorana CP phases) can be determined. A summary of the results for the various cases is given in Table 3. (The results in the case of $\bar M^{}_{e\tau} = 0$ ($\bar M^{}_{\tau\tau} = 0$) can be obtained from those in the case of $\bar M^{}_{e\mu} = 0$ ($\bar M^{}_{\mu\mu} = 0$) by making the replacements $\rho \to \pi -\rho$ and $\sigma \to \pi -\sigma$.) These results will be useful for investigating which, if any, specific texture of neutrino mass matrix is realized by the nature.

We also give some discussions about the model realization of the obtained textures.
It is found that the textures obtained in the case of diagonalization conditions A(G) and C(F) holding automatically in combination with $\bar M^{}_{ee}$, $\bar M^{}_{\mu\tau}$, $m^{}_1$ or $m^{}_3$ being vanishing can reproduce those in the case of $\mu$-$\tau$ reflection symmetry with $M^{}_{ee}$, $M^{}_{\mu\tau}$, $m^{}_1$ or $m^{}_3$ being vanishing, if one restores the unphysical phases with the help of $\bar M^{}_{\alpha \beta} = M^{}_{\alpha \beta} e^{-{\rm i}(\phi^{}_\alpha+ \phi^{}_\beta)}$ and takes $\phi^{}_e = \pi/2$ and $\phi^{}_\mu =  -\phi^{}_\tau$. This in some sense indicates that the approach adopted by us is reasonable. Although the textures obtained in other cases can not be connected to a similar flavor symmetry, they are on an equal footing in our approach and deserve some attention from the phenomenological point of view. Finally, we give some discussions about the breakings of the obtained textures so as to accommodate the deviations of $\theta^{}_{23}$ and $\delta$ from $\pi/4$ and $-\pi/2$. From the example cases we have studied, it is found that small deviations of the physical parameters from their values in the symmetry limit can be accommodated by small breakings of the given textures. And the results obtained in the symmetry limit would still be viable to some extent provided that the given textures hold to a good approximation. Furthermore, because of having distinct main origins, a considerable $\Delta \theta^{}_{23}$ does not necessarily signify a considerable $\Delta \delta$, $\Delta \rho$ or $\Delta \sigma$, and vice versa.

\vspace{0.5cm}

\underline{Acknowledgments} \hspace{0.2cm} We would like to thank the anonymous referees for their valuable comments
and suggestions which help improve our manuscript greatly.
This work is supported
in part by the National Natural Science Foundation of China under grant Nos. 11875157 (Z.C.L. and C.X.Y.), 11605081 (Z.H.Z.) and 11847303.

\end{document}